\documentclass[aps,pr,twocolumn,showpacs,amsmath,amssymb,floatfix,superscriptaddress,reprint,nofootinbib,longbibliography]{revtex4-2}
\usepackage[utf8]{inputenc}
\usepackage[dvips]{graphics}
\usepackage{mathrsfs}
\usepackage[colorlinks=true,citecolor=blue]{hyperref}

\usepackage{subfigure,overpic}
\usepackage{comment}

\newcommand{\up}{\uparrow}
\newcommand{\down}{\downarrow}
\renewcommand{\vec}[1]{\mathbf{#1}}

\providecommand{\nbraket}[1]{\left\langle#1\right\rangle}

\def \beq {\begin{equation}}
\def \edq {\end{equation}}
\def \bes {\begin{subequations}}
\def \eds {\end{subequations}}
\def \beqn {\begin{equation*}}
\def \edqn {\end{equation*}}

\def \dag {\dagger}

\def \up {\uparrow}
\def \down {\downarrow}

\def \sm {\sigma}

\def \veps {\varepsilon}

\def \calh {{\cal{H}}}

\def \calc {{\cal{C}}}

\def \calg {{\cal{G}}}

\def \calt {{\cal{T}}}
\def \calr {{\cal{R}}}

\def \scrg {{\mathscr{G}}}

\def \tGamma {\widetilde{\Gamma}}

\def \ket {\rangle}
\def \bra {\langle}
\def \hc {\text{H.c.}}
\def \veps {\varepsilon}

\begin{document}
\title{Phase-controlled heat modulation with Aharonov-Bohm interferometers}
\author{Sun-Yong Hwang}
\affiliation{Theoretische Physik, Universit\"at Duisburg-Essen and CENIDE, D-47048 Duisburg, Germany}
\author{Björn Sothmann}
\affiliation{Theoretische Physik, Universit\"at Duisburg-Essen and CENIDE, D-47048 Duisburg, Germany}
\author{Rosa L\'opez}
\affiliation{Institut de F\'{\i}sica Interdisciplin\`aria i Sistemes Complexos
IFISC (CSIC-UIB), E-07122 Palma de Mallorca, Spain}
\date{\today}

\begin{abstract}
A heat modulator is proposed based on a voltage-biased Aharonov-Bohm interferometer. Once an electrical bias is applied, Peltier effects give rise to a flow of heat that can be modulated by a magnetic flux. We determine the corresponding temperature changes using a simple thermal model. Our calculations demonstrate that the modulated temperature difference can be as large as 80 mK at base temperature about 600 mK with relative temperature variations reaching 10\%. Our model also predicts, quite generally, the emergence of spin-polarized heat flows without any ferromagnetic contacts, if Rashba spin-orbit interaction is combined with the applied magnetic flux, which potentially paves the way towards caloritronic information processing.
\end{abstract}

\maketitle
\section{Introduction}
Manipulating heat currents and temperatures in mesoscopic devices is still an on-going issues in condensed matter physics community \cite{giazotto_opportunities_2006,dubi_colloquium:_2011,RevModPhys.93.041001}. Heat dissipations at the nanoscale are usually detrimental phenomena that must be better understood and carefully engineered in order to enhance device functionalities.

Recent efforts to deal with unavoidably present waste heat in nanodevices have led to fruitful progress in novel technologies in the newly developed fields such as spin caloritronics \cite{bauer_spin_2012} and phase-coherent caloritronics \cite{fornieri_towards_2017,Hwang2020}, in addition to recently reignited thermoelectrics in the quantum regime \cite{sothmann_thermoelectric_2015,benenti2017fundamental,arrachea2022energy}. In this regards, various caloritronic devices have been proposed that utilize heat in order to construct logical circuits instead of conventional charge degree of freedom. This includes heat interferometers~\cite{giazotto_phase-controlled_2012,giazotto_josephson_2012,martinez-perez_fully_2013,fornieri_nanoscale_2016,fornieri_0-pi_2017}, diodes~\cite{martinez-perez_efficient_2013,giazotto_thermal_2013,fornieri_normal_2014,martinez-perez_rectification_2015}, transistors~\cite{giazotto_proposal_2014,fornieri_negative_2016}, heat switches~\cite{sothmann_high-efficiency_2017}, circulators \cite{hwang_phase-coherent_2018,acciai_phase-coherent_2021}, thermal memory~\cite{guarcello_josephson_2018}, and mesoscopic refrigerators~\cite{solinas_microwave_2016,hofer_autonomous_2016,vischi_coherent_2017}. Central to these phenomena is quantum phase coherence.

Another useful way of thermal managements can be converting electrical work into heat flow via the Peltier effect that can increase or decrease the temperature of reservoirs where the latter case constitutes refrigerators \cite{edwards_a_1993,edwards_cryogenic_1995,prance_electronic_2009,gasparinetti_probing_2011,timofeev_electronic_2009,arrachea_heat_2007,rey_nonadiabatic_2007,cleuren_cooling_2012,
bruggemann_cooling_2014,pekola_refrigerator_2014,nahum_electronic_1994,leivo_efficient_1996,muhonen_micrometre-scale_2012,rouco_electron_2018,linden_how_2010,brunner_virtual_2012,levy_quantum_2012,brunner_entanglement_2014,correa_quantum-enhanced_2014,correa_multistage_2014,venturelli_minimal_2013,entin-wohlman_enhanced_2015,sanchez_correlation-induced_2017,erdman2018absorption,wang_nonlinear_2018,sanchez_cooling_2018,hussein_nonlocal_2019,sanchez_nonlinear_2019,dare2019comparative,hwang_superconductor_2023}.
However, controlling heat flow by electrical means always confronts Joule heating effects leading to unidirectional flow of heat with corresponding temperature increases. Thus, an additional switching knob is neccessary in order to construct heat and temperature modulators such as magnetic flux \cite{fornieri_nanoscale_2016,sothmann_high-efficiency_2017}.

\begin{figure}[b]
\centering
  \begin{centering}
    \includegraphics[width=0.4\textwidth,clip]{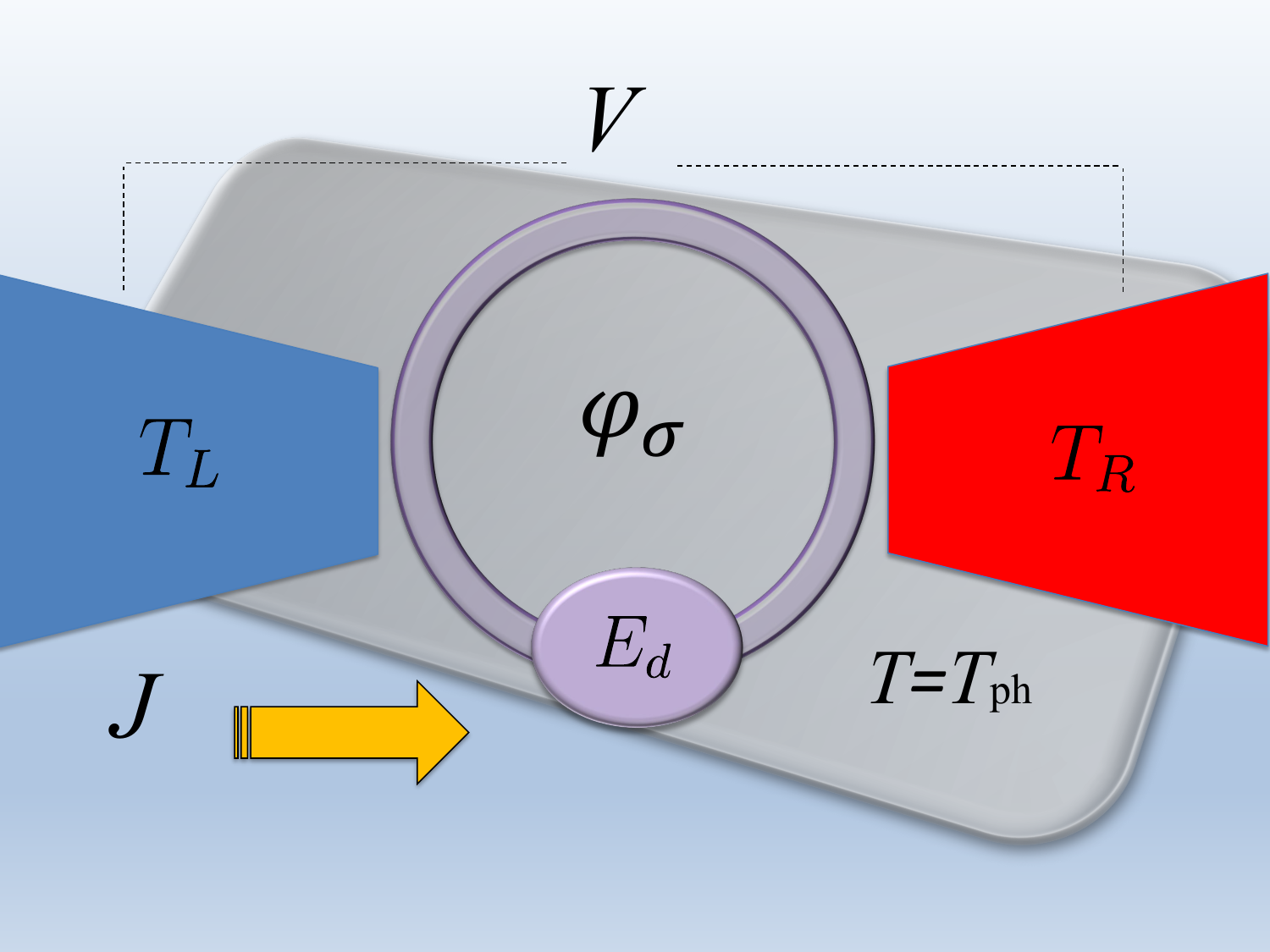}
  \end{centering}
  \caption{Device setup based on voltage-biased Aharonov-Bohm interferometer. A spin-dependent phase $\varphi_\sm$ can appear without any magnetic contacts if both the Ahoronov-Bohm and Rashba phases are present, see Eq. \eqref{phase}. Heat flow $J$ and the temperature of the left reservoir $T_L$ can be modulated periodically with $\varphi_\sm$. We assume that the substrate phonon and the right reservoir share the common background temperature $T=T_{\text{ph}}=T_R$ and apply the thermal model, cf. Eqs. \eqref{eq:el-ph}, \eqref{DT}, \eqref{balance} to determine the temperature change.}
  \label{fig:sketch}
\end{figure}

In this paper, we propose a voltage- and phase-controlled heat modulator based on the Aharonov-Bohm interferometer, see Fig. \ref{fig:sketch}. Such an interferometric device has been studied from the perspective of thermoelectric engines and coolers \cite{kim_thermoelectric_2003,entin-wohlman_three-terminal_2010,entin-wohlman_three-terminal_2012,hwang_proposal_2013,entin-wohlman_enhanced_2015,samuelsson_optimal_2017,haack_efficient_2019,blasi_hybrid_2023} while thermal transport has received no attention so far. 
Furthermore, this type of interferometer has been used to experimentally observe the phase sensitivity \cite{PhysRevLett.74.4047}, Kondo effect \cite{PhysRevLett.90.196601}, partial coherence \cite{PhysRevLett.92.176802}, elastic and inelastic cotunneling \cite{PhysRevLett.98.036805}, and orbital parity symmetry \cite{Debbarma2022}. Therefore, combined with the significant progress in the field of caloritronics \cite{bauer_spin_2012,fornieri_towards_2017,Hwang2020}, our proposal for a coherent heat modulator can be immediately realized with existing experimental technologies.

We calculate electric and heat currents using nonequilibrium Green's function methods \cite{kim_thermoelectric_2003,PhysRevB.71.165310} in a model where one arm of the interferometer is connected to the quantum dot molecule with a vibrational degree of freedom. Such a molecular quantum dot can possess intrinsic spin-orbit interaction \cite{sun_bias_2006,vernek_kondo_2009,lim_kondo_2010} that can be a useful tool for potential spintronic effects. We self-consistently determine the temperature variations at the left reservoir due to electric bias and phase modulation using a phenomenological thermal model \cite{wellstood_hot_1994}. This process includes emergent thermocurrent contributions in every intermediate step accompanying non-stationary thermal gradients until the stationary nonequilibrium states are reached.

The paper is organized as follows. In Sec. \ref{theory}, we detail the device description and theoretical methods. Section \ref{results} shows our main results of the thermal modulator with corresponding discussions. We finally give concluding remarks in Sec. \ref{conclusion}. Appendix \ref{appen1} provides explicit expressions of the self-energy and the lesser Green's function in Eqs. \eqref{eq:EWR0} and \eqref{delep} to make our paper self-contained.

\section{Model}\label{theory}
We consider an Aharonov-Bohm interferometer with a single-level quantum dot embedded in one of the arms. The interferometer is coupled to two electronic reservoirs. In addition, the quantum dot couples to a thermal phononic bath in the substrate, cf. Fig.~\ref{fig:sketch}.
The Hamiltonian for the proposed setup reads \cite{kim_thermoelectric_2003,PhysRevB.71.165310}
\beq\label{ham}
\calh = \calh_C + \calh_D + \calh_T\,,
\edq
where
\beq
\calh_C = \sum_{\ell=L/R,\vec k,\sm} (\veps_{\ell \vec k\sm}-\mu_\ell) c_{\ell \vec k\sm}^{\dag}c_{\ell \vec k\sm}
\edq
 depicts the two normal metal electronic reservoirs with local chemical potential $\mu_\ell$ for the left ($L$) and right ($R$) side with electron energy $\veps_{\ell \vec k\sm}$, momentum $\vec k$, spin $\sm$, while
\beq\label{hm}
\calh_D = \sum_{\sm} E_d d_{\sm}^{\dag}d_{\sm} + \hbar\omega_0 a^{\dag}a + \lambda(a+a^{\dag})\sum_{\sm} d_{\sm}^{\dag}d_{\sm}
\edq
describes an embedded quantum dot with an energy level $E_d$ which is coupled to a phonon bath characterized by an excitation frequency $\omega_0$ and the coupling strength $\lambda$ between the dot and phonon mode. We envisage the quantum dot as a molecular transistor such as carbon nanotubes or C$_{60}$ \cite{park_nanomechanical_2000,leturcq_franck-condon_2009} with vibrational frequencies of the order of THz. These systems possess intrinsic spin-orbit couplings \cite{Minot2004}, hence providing the means to generate a quantum mechanical phase in a purely electric manner in addition to the conventional Aharonov-Bohm phase in the presence of magnetic field, see below. In Eq. \eqref{hm}, $d_\sm$ and $a$ are the corresponding annihilation operators for the dot and the phonon, respectively. 
Finally, in Eq. \eqref{ham}
\beq\label{ht}
\calh_T = \sum_{\ell,\vec k,\sm} (t_{\ell} c_{\ell \vec k\sm}^{\dag}d_{\sm} + \hc) 
+ \sum_{\vec k,\vec p,\sm} (We^{i\varphi_{\sm}}c_{R\vec p\sm}^{\dag}c_{L\vec k\sm} + \hc)
\edq
accounts for tunneling between the dot and reservoirs, where $c_{\ell \vec k\sm}$ is an annihilation operator for electrons with momentum $\vec k$ and spin $\sigma$ in the reservoir $\ell=L/R$. Note that in Eqs. \eqref{hm} and \eqref{ht}, $d_\sm$ and $c_{\ell k\sm}$ are fermionic operators whereas the phonon annilhilation operator $a$ in Eq. \eqref{hm} is a bosonic one. In Eq. \eqref{ht}, $t_{\ell}$ is the tunneling amplitude between the dot and the reservoirs, i.e., lower arm in Fig. \ref{fig:sketch}, and $W$ describes the direct tunneling process between electronic reservoirs, i.e., upper arm in Fig. \ref{fig:sketch}.
The phase factor in Eq. \eqref{ht} can be written as \cite{sun_bias_2006,vernek_kondo_2009,lim_kondo_2010}
\beq\label{phase}
\varphi_{\sm} = \phi+\text{sgn}(\sm)\varphi\,,
\edq
where $\phi=(e/h)\Phi_B$ is the Aharonov-Bohm phase finite under broken time-reversal symmetry with $\Phi_B$ being the magnetic flux. $\varphi= \alpha_R l$ is the Rashba spin-orbit interaction on the dot with $\alpha_R$ and $l$ being respectively the interaction strength and the size of the quantum dot molecule. As we consider a single-level quantum dot, the spin-orbit interaction does not induce spin-flip transitions. Remarkably, Eq. \eqref{phase} suggests that not only our device functionalities can be controlled by two separate phases either by magnetic ($\phi$) or electric ($\varphi$) ways, but it can also be used for potential spin caloritronics devices \cite{bauer_spin_2012} in combination of the two phases even without any ferromagnetic contacts. This spintronic effect originates from an effective Zeeman field induced from the combination of magnetic flux and local Rashba interaction \cite{lim_kondo_2010}.

The spin-resolved charge and heat currents can be respectively calculated from the time evolution of electron number with spin $\sm$, viz. $N_{L\sm}=\sum_{\vec k}c_{L\vec k\sigma}^{\dag}c_{L\vec k\sigma}$, and the energy $\calh_{L\sm} = \sum_{\vec k} \veps_{L\vec k\sm} c_{L\vec k\sm}^{\dag}c_{L\vec k\sm}$ leaving the left reservoir
\begin{align}
&I_\sm=-(ie/\hbar)\bra[\calh,N_{L\sm}]\ket\,,\label{eq:charge}\\
\label{eq:Jsigma}&J_\sm=-(i/\hbar)\bra[\calh,\calh_{L\sm}]\ket-I_\sm (V_L-V_R)\,,
\end{align}
where $I_\sm (V_L-V_R)$ corresponds to the Joule heating term. 
Eqs. \eqref{eq:charge}, and   \eqref{eq:Jsigma} yield via nonequilibirum Green's function methods \cite{kim_thermoelectric_2003,haug_quantum_2008,PhysRevB.81.155323}
\beq\label{currents}
\begin{pmatrix}
I_{\sm} \\
J_{\sm}
\end{pmatrix}
=
\frac{1}{h}\int d\veps~\begin{pmatrix} e \\ (\veps-\mu) \end{pmatrix}\calt_{\sm}(\veps) \left[f_L(\veps) - f_R(\veps)\right]\,,
\edq
where $\mu=eV+E_F$ and $f_{L}(\varepsilon)=\{1+\exp[(\varepsilon-eV-E_F)/k_{B}T_{L}]\}^{-1}$ is the Fermi distribution function at the left lead with a local temperature $T_L$ whereas $f_R(\varepsilon)=\{1+\exp[(\varepsilon-	E_F)/k_{B}T]\}^{-1}$ is that for the right reservoir. We take the Fermi level $E_F=0$ and the bias voltage configuration $V=V_L$ and $V_R=0$. We assume the phonon and the right reservoir share the same background temperature $T=T_{\text{ph}}=T_R$. The spin-resolved transmission function in Eq. \eqref{currents} is explicitly given by \cite{kim_thermoelectric_2003,hwang_proposal_2013}
\begin{multline}
\calt_{\sm}(\veps)=
\calt_{b\sm}\big\{\big(1+2\tGamma_{\sm}\text{Im}\left[\calg_{\sm,\sm}^r\right]\big) \\
+ \tGamma_{\sm}^2\left[1 - \alpha_{\sm}\cos^2(\varphi_{\sm})\right]\left|\calg_{\sm,\sm}^r\right|^2\big\}+ \tGamma_{\sm}^2 \alpha_{\sm}\left|\calg_{\sm,\sm}^r\right|^2\\
+ 2\tGamma_{\sm}\sqrt{\alpha_{\sm}\calt_{b\sm}(1-\calt_{b\sm})}\cos(\varphi_{\sm})\text{Re}\left[\calg_{\sm,\sm}^r\right] \,,
\label{trans}
\end{multline}
where $\calt_{b\sm}(\veps) =4\xi_{\sm}(\veps)/[1+\xi_{\sm}(\veps)]^2$, $\xi_{\sm}(\veps) = \pi^2\rho_{L\sm}(\veps)\rho_{R\sm}(\veps)W^2$, $\alpha_{\sm}(\veps) = 4\Gamma_{L\sm}\Gamma_{R\sm}/(\Gamma_{L\sm}+\Gamma_{R\sm})^2$, $\tGamma_{\sm}(\veps) = (\Gamma_{L\sm}+\Gamma_{R\sm})/(1+\xi_{\sm}(\veps))$, $\Gamma_{\ell\sm}(\veps) = \pi\rho_{\ell\sm}(\veps)t_{\ell}^2$ with $\rho_{\ell\sm}(\veps)=\sum_{\bf k}\delta(\veps-\veps_{\ell{\bf k}\sm})$. For definiteness, we neglect the spin-polarization of the contacts such that $\Gamma_{\ell\sm}=\Gamma_\ell$ and take the semi-elliptic band for the density of states, i.e., $\rho_{\ell\sm}(\veps)=\rho_{\ell}(\veps)\propto\sqrt{1-(\veps/D)^2}$, with which one writes $\xi_{\sm}(\veps)=\xi(\veps)=\xi_0[1-(\veps/D)^2]$. Thus, an energy-independent dimensionless quantity $\xi_0$ represents a direct tunneling rate only through the upper arm in Fig. \ref{fig:sketch}. For numerical calculations, we fix the half-bandwidth $D=10k_BT$. We remark that the choice of a different band, e.g., Lorentzian $1/[1+(\veps/D)^2]$, does not qualitatively modify the results in our work. In Eq.\eqref{trans}, the retarded Green's function can be written by \cite{entin-wohlman_three-terminal_2012}
\beq
\calg_{\sm,\sm}^{r}(\veps)
= \left[\veps-E_d - \Sigma_{0\sm}^{r}(\veps) - \delta\veps_P - \Sigma_{P\sm}^{r}(\veps) \right]^{-1}\,, 
\label{eq:EWR0}
\edq
exact up to the second order in electron-phonon coupling $\lambda$, where $\Sigma_{0\sm}^{r}(\veps)= -i\tGamma_{\sm} - \tGamma_{\sm}\sqrt{\alpha_{\sm}\xi_{\sm}}\cos(\varphi_\sm)$ is the self-energy due to couplings to the leads. $\delta\veps_P$ and $\Sigma_{P\sm}^{r}(\veps)$ are the energy shift and the self-energy correction arising from the electron-phonon coupling, respectively. The explicity expression for $\delta\veps_P$ reads
\beq\label{delep}
\delta\veps_P = -\frac{2\lambda^2}{\hbar\omega_0} \sum_{\sm}\frac{1}{2\pi i}\int d\veps~\scrg_{\sm,\sm}^<(\veps)\,,
\edq
whereas those of $\Sigma_{P\sm}^{r}(\veps)$ in Eq. \eqref{eq:EWR0} and the lesser Green's function $\scrg_{\sm,\sm}^<(\veps)$ in Eq. \eqref{delep} are provided in Appendix \ref{appen1}. As one can easily notice, the effects of $\delta\veps_P$ and $\Sigma_{P\sm}^{r}(\veps)$ are the renormalization of the dot level $E_d\to E_d+\delta\veps_P$ and the tunnel couplings $\Sigma_{0\sm}^{r}(\veps)\to \Sigma_{0\sm}^{r}(\veps)+ \Sigma_{P\sm}^{r}(\veps)$, respectively. We focus on a noninteracting quantum dot, as our primary interest lies in the interference effects that are revealed through phase dependence. Furthermore, at the mean-field level, interaction only yields a shift in the energy level, hence effectively playing the same role as $\delta\veps_P$ quantitatively. Once the charge and spin currents are determined via Eq. \eqref{eq:charge} as $I_c=I_\up+I_\down$ and $I_s = I_{\up} - I_{\down}$, one can then evaluate the total heat flux $J$ and the spin-polarized heat $J_s$ via Eq. \eqref{eq:Jsigma}, i.e.,
\begin{align}
J&=J_\up+J_\down\,,\label{J}\\
J_s&=J_\up-J_\down\,,\label{Js}
\end{align}
that are of our main interest and can be controlled by phases $\phi$ or $\varphi$, cf. Eq. \eqref{phase}.

In Eq. \eqref{currents}, we have only considered the elastic contribution to the currents although there exists an inelastic term of the order $\lambda^2$ containing the phase-dependent term proportional to $\sin\varphi_\sm$ \cite{PhysRevB.81.155323}. We justify this by taking the small value of $\lambda=0.1k_BT$ in the numerical calculations. Further justification can be made by assuming that this inelastic term can be safely absorbed into the phenomenological equation based on the thermal model, where the heat flow from the substrate can be written by \cite{wellstood_hot_1994}
\beq\label{eq:el-ph}
P_{\text{el-ph}}=\Sigma{\mathcal{V}}(T_L^5-T_\text{ph}^5)\,,
\edq
where $\Sigma$ and ${\mathcal{V}}$ respectively denote the strength of electron-phonon coupling and the volume of the device. We take $\Sigma\simeq10^9$~WK$^{-5}$m$^{-3}$ and ${\mathcal{V}}\simeq10^{-20}$m$^{3}$, that are realistic values for a typical mesoscopic setup \cite{giazotto_opportunities_2006}. Initially, we consider an isothermal situation $T_L=T=T_R$, and since the applied voltage generates the heat flow due to Peltier effect, the temperature of the left lead also changes
\beq\label{DT}
T_L=T+\delta T\,,
\edq
where the amount of change $\delta T$ can be determined via Eqs. \eqref{J} and \eqref{eq:el-ph} from the heat balance equation
\beq\label{balance}
J+P_{\text{el-ph}}=0\,.
\edq
We self-consistently include the emerging thermocurrents whenever the temperature gradient is generated until the stationary solutions of the heat flow $J$ and $\delta T$ are reached \cite{hwang_superconductor_2023}. We will not consider applied thermal gradient across the system, thus the control of heat and temperature in our device are managed with electrical voltages. We also mention that if the dissipative Joule heating contribution becomes large one should in general consider the temperature change of both reservoirs while we have only considered that of left metal for definiteness, following Ref. \cite{hwang_superconductor_2023}. Furthermore, the case with strong Coulomb interactions at the quantum dot with an arbitrary strength of the electron-phonon coupling is beyond the scope of this work and hence left as future investigations.

\begin{figure}[t]
\centering
\includegraphics[width=0.4\textwidth]{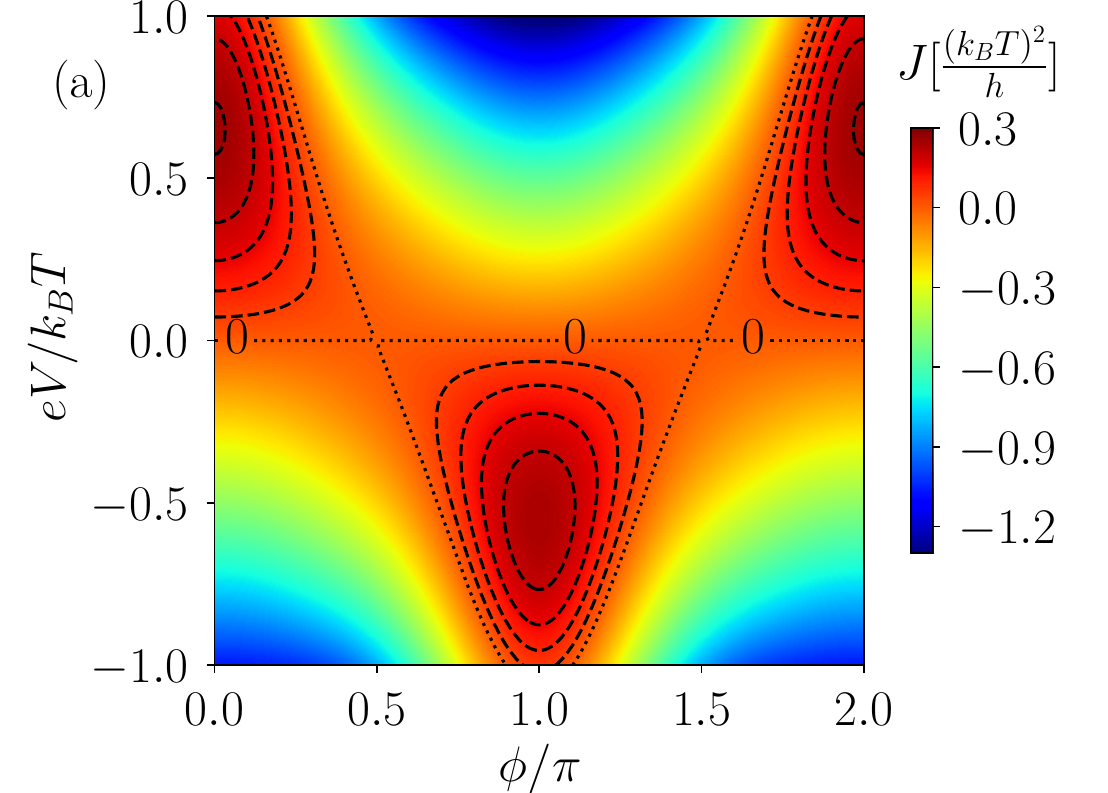}\\
\includegraphics[width=0.4\textwidth]{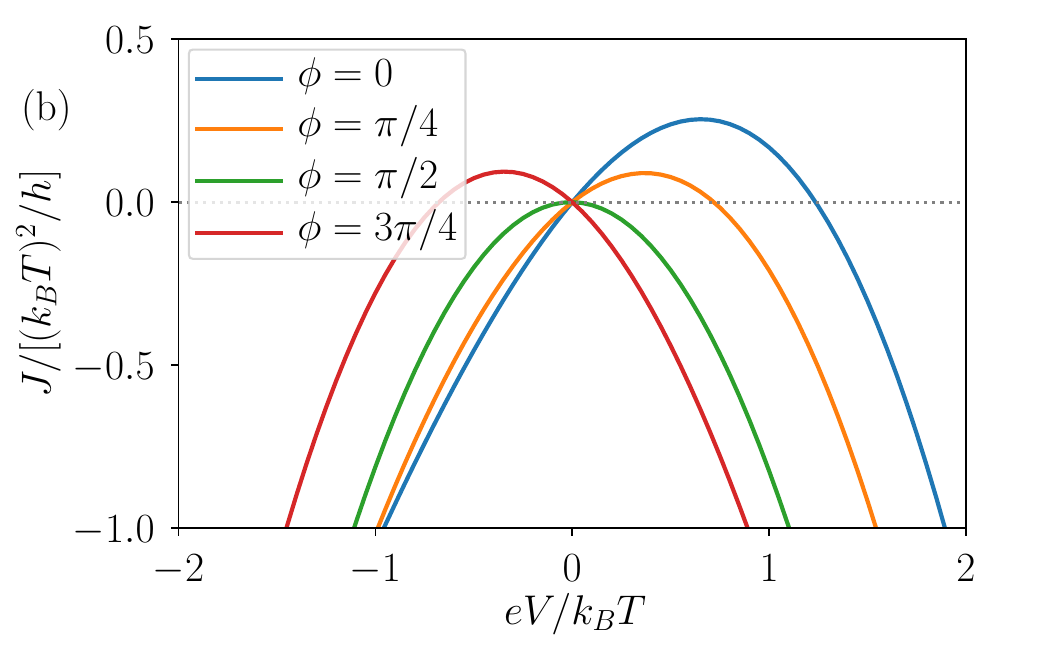}\\
\includegraphics[width=0.4\textwidth]{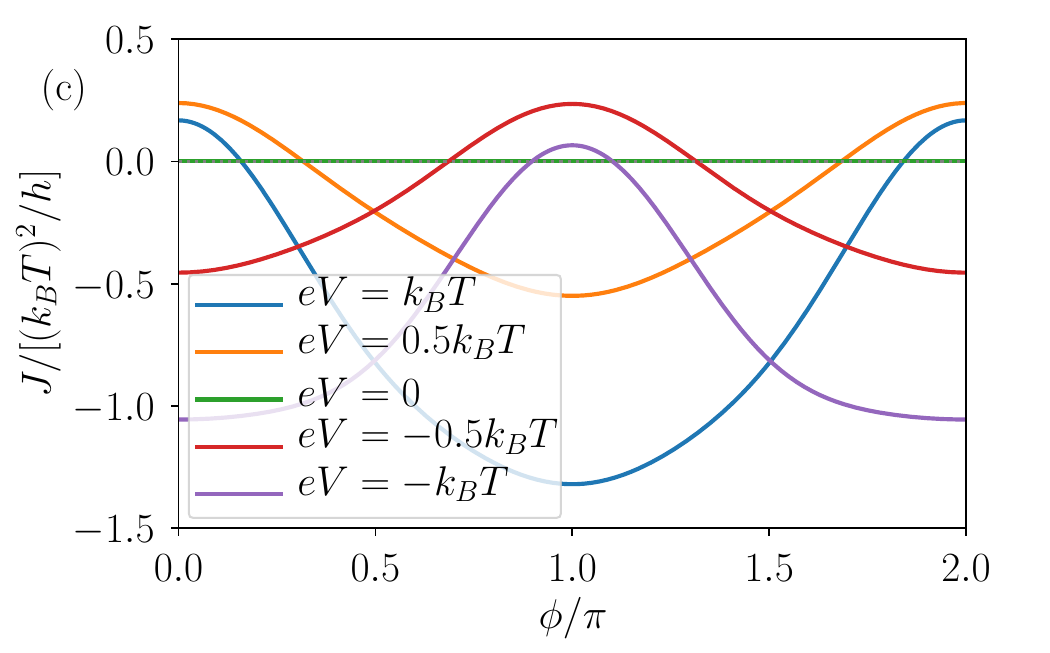}
  \caption{(a) $J$ versus $\phi$ and $eV$. Dotted lines represent a vanishing heat current $J = 0$, while the dashed lines indicate the region where the left reservoir is cooled down, i.e.,  $J > 0$. (b) and (c) depict cross-sectional views of (a) along the vertical ($\phi=0,\frac{\pi}{4},\frac{\pi}{2},\frac{3\pi}{4}$) and horizontal axes ($\frac{eV}{k_BT}=1,0.5,0,-0.5,-1$), respectively. Rest of parameters: $E_d=0.3k_BT$, $\hbar\omega_0=0.2k_BT$, $\lambda=0.1k_BT$, $\Gamma_L=\Gamma_R=0.5k_BT$, and $\xi_0=0.4$, $\varphi=0$ (mod $2\pi$).}
  \label{Fig2}
\end{figure}

\section{Results and discussion}\label{results}
Figure  \ref{Fig2}(a) displays the contour map of the steady heat flow $J$ as functions of an Aharonov-Bohm (AB) phase ($\phi$) and an applied voltage ($V$), in units of $(k_BT)^2/h$. The dotted lines indicate vanishing heat current ($J=0$) and the dashed lines denote the region where net cooling of the left reservoir can be achieved, i.e., $J>0$. Notably, the sign of $J$ can be easily controlled by varying the magnetic flux ($\Phi_B\propto\phi$) threaded through the device at a fixed voltage bias. As indicated in Eq. \eqref{phase}, the role of $\phi$ can be replaced with the Rashba phase $\varphi$ since an elastic contribution of transmission function only contains terms proportional to $\cos\varphi_\sm$ hence the role of two phases can be interchanged in the absence of one another. Practically, however, controlling with the magnetic flux will be easier to vary the effects of a phase than by tuning the spin-orbit interaction.

In Fig. \ref{Fig2}(b), $J$ is plotted as a function of voltage for several given AB phases. $J$ can be nonzero as the voltage is applied and can become positive, i.e., net cooling effect, until the Joule heating dominates making $J$ with large negative values. Nevertheless, one can notice that at a given voltage, the sign of $J$ can be easily switched with varying $\phi$. This is more clearly visible in Fig. \ref{Fig2}(c) where we plot it as a function of $\phi$ for several chosen values of voltage. The direction of heat flow can be conveniently modulated with AB phase as a control knob. It should be noted that there are optimum ranges of $V$ within which periodic sign change of heat is more pronounced since large voltages finally hinder cooling-heating modulation effects invisible due to Joule heating, e.g., $|eV|=k_BT$ in Fig. \ref{Fig2}(c). Nevertheless, even in full heating regime characterized by $J<0$, the amplitude of heat modulation is visibly large which thus induces a large temperature modulation in our device.

\begin{figure*}[t]
\centering
  \begin{tabular}{cc}
    \includegraphics[width=0.43\textwidth]{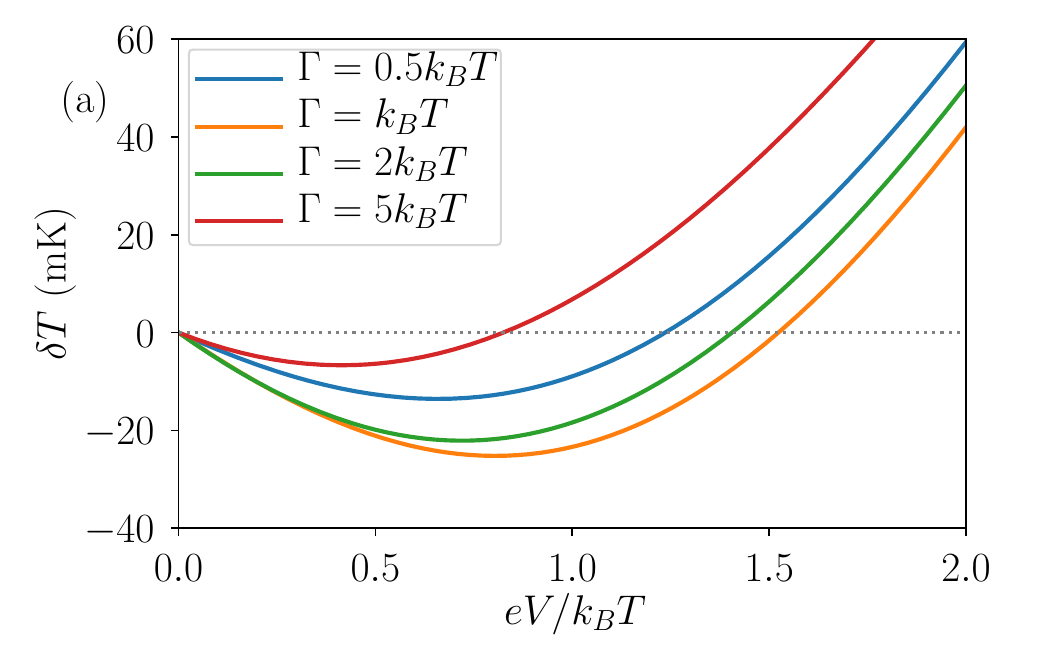}&
 \includegraphics[width=0.35\textwidth]{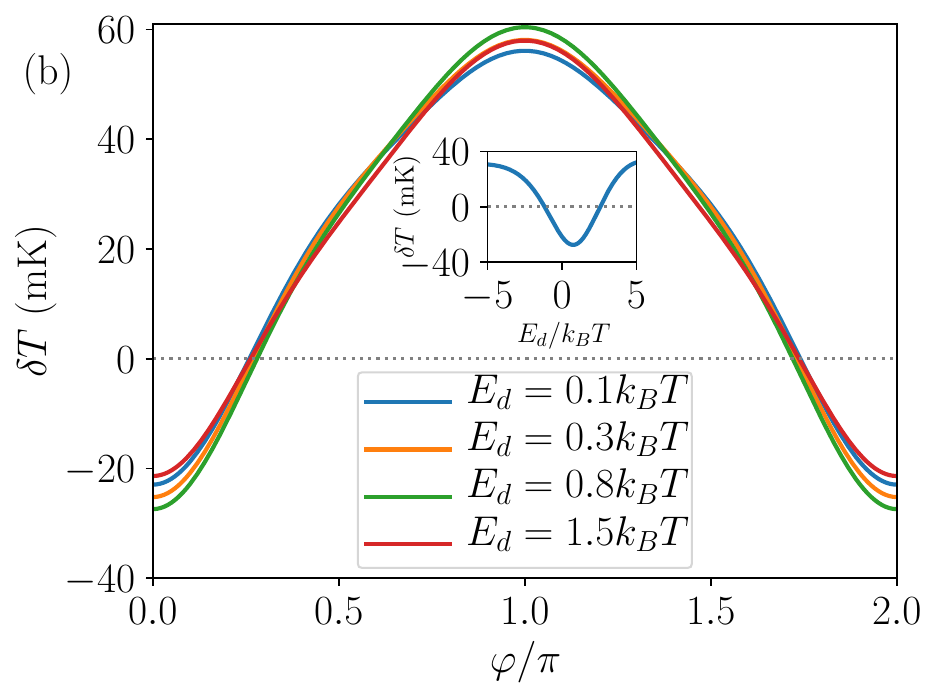}\\
 \includegraphics[width=0.43\textwidth]{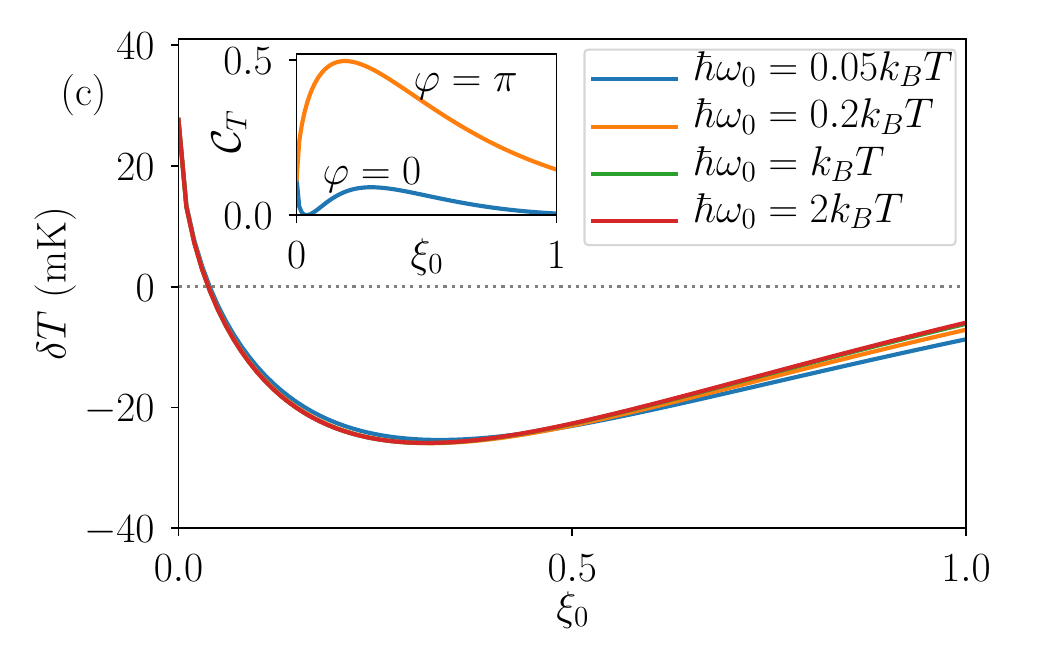}&
\includegraphics[width=0.43\textwidth]{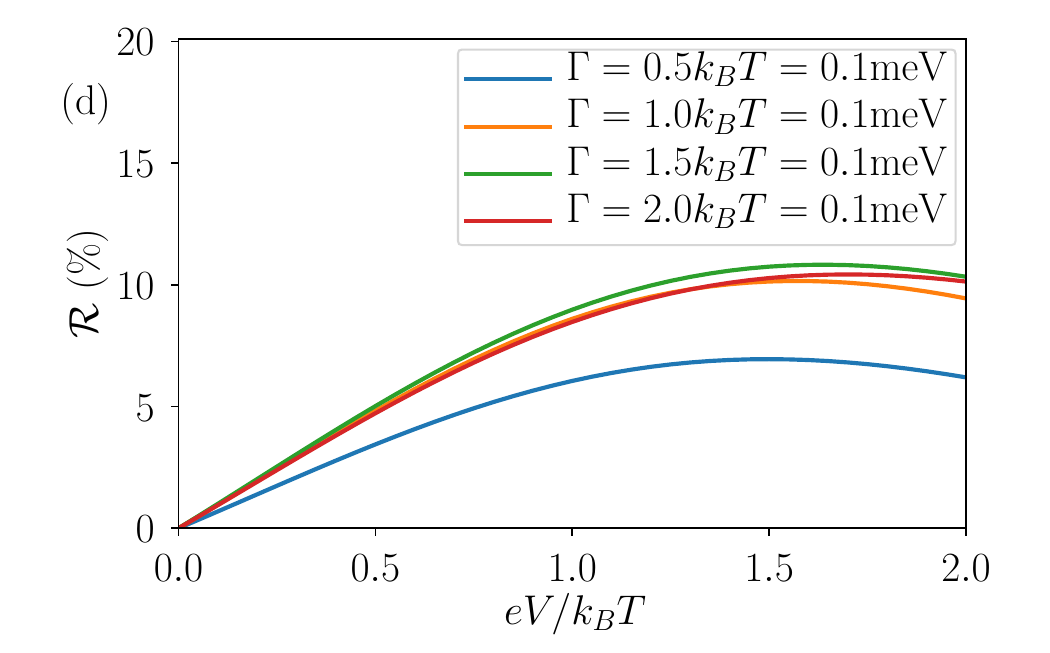}
  \end{tabular}
  \caption{$\delta T$ versus (a) $eV$ for various $\Gamma=\Gamma_{L}=\Gamma_R$ at $\phi=\varphi=0$, $E_d=0.3k_BT$, $\hbar\omega_0=0.2k_BT$, (b) $\varphi$ for several $E_d$ at $\phi=0$, $eV=0.8k_BT$ and $\Gamma=k_BT$, (c) $\xi_0$ for several $\omega_0$ at $\phi=\varphi=0$, $E_d=0.3k_BT$, $eV=0.8k_BT$ and $\Gamma=k_BT$ with $T\approx580$ mK, (d) $\calr$ vs. $V$ for various $T$ at a fixed $\Gamma=0.1$ meV. In (a),(b),(d), we take $\xi_0=0.4$ and in all cases, we set $\lambda=0.1k_BT$. Inset in (b) is a plot as a function of $E_d$ at $\varphi=0$. Inset in (c) shows $\calc_T$ vs $\xi_0$ at $\hbar\omega_0=0.2k_BT$ for $\varphi=0$ and $\varphi=\pi$ while other parameters are fixed.}
  \label{Fig3}
\end{figure*}

We explore now the dependence of temperature modulation $\delta T$ cf. Eq. \eqref{DT}, on several model parameters such as the tunnel broadening $\Gamma$ and the phonon frequency $\omega_0$ as the voltage and the phase are varied. Figure \ref{Fig3} shows such dependences of $\delta T$. In particular, Fig. \ref{Fig3}(a) shows the accompanying temperature modulation $\delta T$, plotted as a function of voltage bias at fixed phases $\phi=\varphi=0$ (mod $2\pi$) for selected values of coupling strengths $\Gamma$. Unsurprisingly, within a certain range of voltages there exist cooling effects $\delta T<0$ after which Joule heating largely dominates and hence the temperature increases producing a positive $\delta T>0$. However, at a fixed $V$ as shown in Fig. \ref{Fig3}(b) with $eV=0.8k_BT$, $\delta T$ can be periodically modulated with a phase either by $\phi$ or $\varphi$ although only the latter example is displayed here. Remarkably, temperature difference between the maximum at $\varphi=\pi$ and the minimum $\varphi=2n\pi$ ($n$: integer) can be as large as 80 mK, which is much larger than the experimentally achievable temperature measurement resolution of the order of 0.1 mK {\cite{giazotto_opportunities_2006}}. Furthermore, modulations are robust against unintended variations of $E_d$ as can be seen from Fig. \ref{Fig3}(b), or put it another way, do not resort to a fine tuning of $E_d$ establishing the shown phenomena mainly from phase effects. The inset of Fig. \ref{Fig3}(b) displays effects of a large variations of $E_d$ at $\varphi=0$ while other parameters fixed. It is shown that for a refrigeration effect to occur at $\varphi=2n\pi$, there exists a broad range of $E_d$ where $\delta T<0$. For larger values of $|E_d|$, the temperature finally rises. 
Figure \ref{Fig3}(c) shows the effects of varying background transmission $\xi_0$ for different values of the phonon frequency at a fixed phase. It is not surprising that $\delta T$ is rather insensitive to the variation of $\omega_0$ since in our model of weak electron-phonon coupling, the main effect of phonons is a renormalization of the level position and the coupling strengths, cf. Eq. \eqref{eq:EWR0}. For $\xi_0\to0$, i.e., in the limit where the upper arm is absent in Fig. \ref{fig:sketch}, the sign of $\delta T$ can change from negative to positive indicating the role of interference for refrigeration effects in this parameter regime. In the inset of Fig. \ref{Fig3}(c), the coefficient of thermodynamic uncertainty relation violation \cite{agarwalla_assessing_2018,kheradsoud_power_2019} $\calc_T$ is plotted at $\hbar\omega_0=0.2k_BT$ as $\xi_0$ is varied, where
\beq
\calc_T=2\frac{J^2}{S_J}\cdot\frac{k_B}{\sigma_E}\,,
\edq
with the heat current noise given by
\begin{multline}
S_J=\frac{1}{h}\int d\veps(\veps-eV)^2\Big\{\sum_\sm\calt_\sm(\veps)\sum_{\ell=L,R}[f_\ell(\veps)(1-f_\ell(\veps)]\\
+\sum_\sm\calt_\sm(\veps)[1-\sum_\sm\calt_\sm(\veps)][f_L(\veps)-f_R(\veps)]^2\Big\}\,,
\end{multline}
and the entropy production being $\sigma_E=IV/T+J\delta T/T^2$. As shown, the thermodynamic uncertainty relation is not violated, i.e., $\calc_T\le1$, in our system. However, $\calc_T$ becomes maximal, i.e., closer to the violation, roughly when the cooling effects become the largest with $\delta T<0$, indicating that fluctuations generate less entropy in the refrigeration regime.

\begin{figure}[t]
\centering
\includegraphics[width=0.38\textwidth]{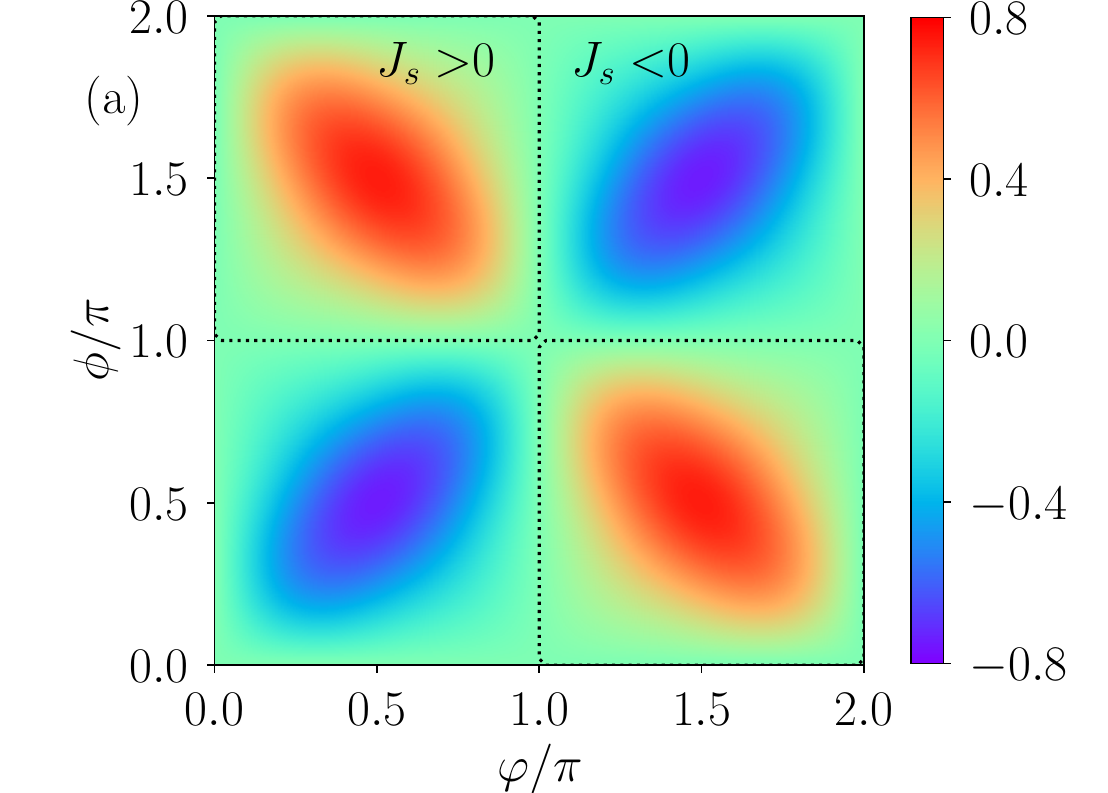}\\
\includegraphics[width=0.38\textwidth]{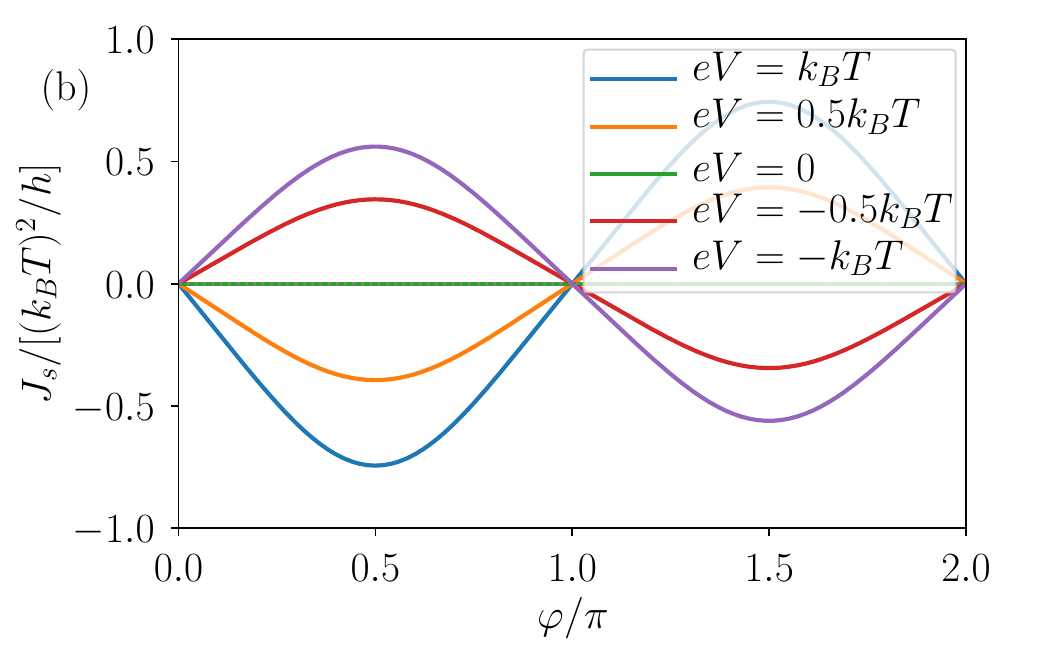}
  \caption{$J_s$ versus (a) $\varphi$ and $\phi$ at $eV=k_BT$, (b) $\varphi$ for several $eV$ at $\phi=\pi/2$. Parameters are $E_d=0.3k_BT$, $\hbar\omega_0=0.2k_BT$, $\lambda=0.1k_BT$, $\Gamma_L=\Gamma_R=0.5k_BT$, and $\xi_0=0.4$. Dotted lines indicate the boundary where $J_s=0$.}
  \label{Fig4}
\end{figure}

In order to quantify the figure of merit for the temperature modulation in our device, we introduce a relative temperature variation defined by \cite{fornieri_nanoscale_2016,sothmann_high-efficiency_2017}
\beq
\calr=\frac{T_L^{\text{max}}-T_L^{\text{min}}}{T_L^{\text{min}}}\,,
\edq
where $T_L^{\text{max}}$ ($T_L^{\text{min}}$) is the maximum (minimum) temperature obtained by the phase modulation, e.g., at $\varphi=\pi$ ($\varphi=2n\pi$) in Fig. \ref{Fig3}(b). $\calr$ is plotted in Fig. \ref{Fig3}(d) as $V$ is applied for different base temperatures ranged approximately from 500 mK to 1 K, where one could get $\calr\sim10\%$ for optimum conditions. This value is comparable in magnitude to experimentally obtained values \cite{fornieri_nanoscale_2016} as well as a theoretical estimation of high-efficiency thermal switch based on the topological Josepshon junction \cite{sothmann_high-efficiency_2017}, with an added advantage in our setup that we do not need a SQUID geometry with superconducting Josephson junctions.

Finally, we discuss about the spin-polarized heat flow, cf. Eq. \eqref{Js}, which may be relevant to recent developments in spin caloritronics \cite{bauer_spin_2012} and phase-coherent caloritronics \cite{fornieri_towards_2017,Hwang2020} that envisage novel quantum information processing devices based on the generated heat flow. In Fig. \ref{Fig4}(a), $J_s$ is shown as functions of two phases that can be respectivly controlled magnetically and electrically. As discussed above, one can notice that $J_s$ is symmetric when the role of the magnetic AB phase and the electric Rashba phase are interchanged, cf. Eq. \eqref{phase}, since the cosine function is even and the transmission in Eq. \eqref{trans} contains terms including only $\cos\varphi_\sm$. Importantly, for $J_s$ to be nonzero, both phases should be present thus an essential requirement for spintronic effects in the device is a finite magnetic field. Otherwise, everything else involves purely electrical means without any ferromagnetic contacts.
Remarkably, both for $J_s>0$ and $J_s<0$ regimes, corresponding powers are almost equal in strength in stark contrast to the total heat flux $J$ as shown in Fig. \ref{Fig2}(a). This can be more visible in Fig. \ref{Fig4}(b) where $J_s$ is plotted as a function of Rashba phase for a fixed $\phi=\pi/2$. As shown, the amplitude and the sign of $J_s$ signals can be controlled electrically by applying voltage bias across the system for fixed phases.

\section{Conclusions}\label{conclusion}
We have proposed a heat and temperature modulator based on the mesoscopic Aharonov-Bohm intereferometer which is voltage biased. The direction and the amplitude of the heat flow can be easily controlled by either Aharonov-Bohm or Rashba phase with the aid of electrical bias. Correspondingly, temperature variations can be also modulated by the phases working as control knobs. The temperature variations is of  the order of 80 mK with accompanying heat flow of the order of hundreds fW. The relative temperature variations can reach 10\% which is comparable to reported values in experiments and other theoretical proposals. Our proposal provides an easy manipulation of the heat flow and the temperature variations, both the direction and the amplitude, exploiting the Aharonov-Bohm and Rashba phases.
Furthermore, in general cases with both phases present, our device can generate spin-polarized heat flows even without any ferromagnetic leads, that can switch the sign easily by the magnetic flux and the voltage bias. Potentially, this spintronic heat flow can be used for information processing devices in the caloritronic setup. Thus, our results are strongly relevant to recently proposed novel technologies such as spin and phase-coherent caloritronics.

\begin{acknowledgments}
SYH is so much grateful to J. Nied\'zwiedzka and W. Wnuczy\'nska for insightful discussions.
We acknowledge financial support from the Ministry of Innovation NRW via the ``Programm zur Förderung der Rückkehr des hochqualifizierten Forschungsnachwuchses aus dem Ausland''.
R.L. acknowledges the financial support by Grant No.PDR2020/12 sponsored by the Comunitat Autonoma de les Illes Balears through the Direcció General de Política Universitaria i Recerca with funds from the Tourist Stay Tax
Law ITS 2017-006, Grants No. PID2020-117347GB-I00, and No. LINKB20072 from the CSIC i-link program 2021.
This work has been partially supported by the María de Maeztu project CEX2021-001164-M funded by the MCIN/AEI/10.13039/501100011033.
\end{acknowledgments}

\appendix
\section{Lesser Green's function and the retarded phonon self-energy}\label{appen1}
The lesser (greater) Green's function in Eq. \eqref{delep} is given by
\beq
\scrg_{\sm,\sm}^{</>}(\veps) = \scrg_{\sm,\sm}^r(\veps) \Sigma_{0\sm}^{</>}(\veps) \scrg_{\sm,\sm}^{r,*}(\veps)\,,
\edq
where $\scrg_{\sm,\sm}^{r}(\veps) = [\veps-E_d+i\tGamma_{\sm} +\tGamma_{\sm}\sqrt{\alpha_{\sm}\xi_{\sm}}\cos(\varphi_{\sm})]^{-1}$,
\begin{multline}
\Sigma_{0\sm}^< (\veps)= \frac{2i}{(1+\xi_{\sm})^2}[\left(\Gamma_{L\sm}+\xi_{\sm}\Gamma_{R\sm}\right)f_{L\sm}(\veps) \\
+ \left(\Gamma_{R\sm}+\xi_{\sm}\Gamma_{L\sm}\right)f_{R\sm}(\veps)]\\
+ \frac{i\Gamma_{\sm}}{1+\xi_{\sm}}\sqrt{\alpha_{\sm}\calt_{b\sm}}\sin(\varphi_{\sm})(f_{L\sm}(\veps)-f_{R\sm}(\veps))\,,
\end{multline}
and
\begin{multline}
\Sigma_{0\sm}^>(\veps) 
= -\frac{2i}{(1+\xi_{\sm})^2}[\left(\Gamma_{L\sm}+\xi_{\sm}\Gamma_{R\sm}\right)(1-f_{L\sm}(\veps)) \\
+ \left(\Gamma_{R\sm} +\xi_{\sm}\Gamma_{L\sm}\right)(1-f_{R\sm}(\veps))]\\
+ \frac{i\Gamma_{\sm}}{1+\xi_{\sm}}\sqrt{\alpha_{\sm}\calt_{b\sm}}\sin(\varphi_{\sm})(f_{L\sm}(\veps)-f_{R\sm}(\veps))\,.
\end{multline}

Equipped with lesser and greater Green's functions, the self-energy correction due to phonon in Eq. \eqref{eq:EWR0} can be written as
\begin{multline}
\Sigma_{P\sm}^r(\veps)
= i\lambda^2\int \frac{d\veps'}{2\pi}~
\bigg\{\bigg(\frac{\nbraket{aa^{\dag}}}{\veps - \hbar\omega_0 - \veps' + i0^+}  \\ 
+ \frac{\nbraket{a^{\dag}a}}{\veps + \hbar\omega_0 - \veps' + i0^+}\bigg)\scrg_{\sm,\sm}^>(\veps') \\
- \bigg(\frac{\nbraket{aa^{\dag}}}{\veps + \hbar\omega_0 - \veps' + i0^+} + \frac{\nbraket{a^{\dag}a}}{\veps - \hbar\omega_0 - \veps' + i0^+}\bigg)\scrg_{\sm,\sm}^<(\veps')\bigg\}\,,
\end{multline}
where $\nbraket{a^{\dag}a}=\nbraket{aa^{\dag}}-1=N_{ph}= (e^{\hbar\omega_0/k_BT_{\text{ph}}}-1)^{-1}$ is the Bose-Einstein occupation for the phonon mode.

\bibliography{ABcooler}

\begin{thebibliography}{81}%
\makeatletter
\providecommand \@ifxundefined [1]{%
 \@ifx{#1\undefined}
}%
\providecommand \@ifnum [1]{%
 \ifnum #1\expandafter \@firstoftwo
 \else \expandafter \@secondoftwo
 \fi
}%
\providecommand \@ifx [1]{%
 \ifx #1\expandafter \@firstoftwo
 \else \expandafter \@secondoftwo
 \fi
}%
\providecommand \natexlab [1]{#1}%
\providecommand \enquote  [1]{``#1''}%
\providecommand \bibnamefont  [1]{#1}%
\providecommand \bibfnamefont [1]{#1}%
\providecommand \citenamefont [1]{#1}%
\providecommand \href@noop [0]{\@secondoftwo}%
\providecommand \href [0]{\begingroup \@sanitize@url \@href}%
\providecommand \@href[1]{\@@startlink{#1}\@@href}%
\providecommand \@@href[1]{\endgroup#1\@@endlink}%
\providecommand \@sanitize@url [0]{\catcode `\\12\catcode `\$12\catcode
  `\&12\catcode `\#12\catcode `\^12\catcode `\_12\catcode `\%12\relax}%
\providecommand \@@startlink[1]{}%
\providecommand \@@endlink[0]{}%
\providecommand \url  [0]{\begingroup\@sanitize@url \@url }%
\providecommand \@url [1]{\endgroup\@href {#1}{\urlprefix }}%
\providecommand \urlprefix  [0]{URL }%
\providecommand \Eprint [0]{\href }%
\providecommand \doibase [0]{https://doi.org/}%
\providecommand \selectlanguage [0]{\@gobble}%
\providecommand \bibinfo  [0]{\@secondoftwo}%
\providecommand \bibfield  [0]{\@secondoftwo}%
\providecommand \translation [1]{[#1]}%
\providecommand \BibitemOpen [0]{}%
\providecommand \bibitemStop [0]{}%
\providecommand \bibitemNoStop [0]{.\EOS\space}%
\providecommand \EOS [0]{\spacefactor3000\relax}%
\providecommand \BibitemShut  [1]{\csname bibitem#1\endcsname}%
\let\auto@bib@innerbib\@empty
\bibitem [{\citenamefont {Giazotto}\ \emph {et~al.}(2006)\citenamefont
  {Giazotto}, \citenamefont {Heikkil\"a}, \citenamefont {Luukanen},
  \citenamefont {Savin},\ and\ \citenamefont
  {Pekola}}]{giazotto_opportunities_2006}%
  \BibitemOpen
  \bibfield  {author} {\bibinfo {author} {\bibfnamefont {F.}~\bibnamefont
  {Giazotto}}, \bibinfo {author} {\bibfnamefont {T.~T.}\ \bibnamefont
  {Heikkil\"a}}, \bibinfo {author} {\bibfnamefont {A.}~\bibnamefont
  {Luukanen}}, \bibinfo {author} {\bibfnamefont {A.~M.}\ \bibnamefont
  {Savin}},\ and\ \bibinfo {author} {\bibfnamefont {J.~P.}\ \bibnamefont
  {Pekola}},\ }\bibfield  {title} {\bibinfo {title} {Opportunities for
  mesoscopics in thermometry and refrigeration: {Physics} and applications},\
  }\href {https://doi.org/10.1103/RevModPhys.78.217} {\bibfield  {journal}
  {\bibinfo  {journal} {Rev. Mod. Phys.}\ }\textbf {\bibinfo {volume} {78}},\
  \bibinfo {pages} {217} (\bibinfo {year} {2006})}\BibitemShut {NoStop}%
\bibitem [{\citenamefont {Dubi}\ and\ \citenamefont
  {Di~Ventra}(2011)}]{dubi_colloquium:_2011}%
  \BibitemOpen
  \bibfield  {author} {\bibinfo {author} {\bibfnamefont {Y.}~\bibnamefont
  {Dubi}}\ and\ \bibinfo {author} {\bibfnamefont {M.}~\bibnamefont
  {Di~Ventra}},\ }\bibfield  {title} {\bibinfo {title} {Colloquium: {Heat} flow
  and thermoelectricity in atomic and molecular junctions},\ }\href
  {https://doi.org/10.1103/RevModPhys.83.131} {\bibfield  {journal} {\bibinfo
  {journal} {Rev. Mod. Phys.}\ }\textbf {\bibinfo {volume} {83}},\ \bibinfo
  {pages} {131} (\bibinfo {year} {2011})}\BibitemShut {NoStop}%
\bibitem [{\citenamefont {Pekola}\ and\ \citenamefont
  {Karimi}(2021)}]{RevModPhys.93.041001}%
  \BibitemOpen
  \bibfield  {author} {\bibinfo {author} {\bibfnamefont {J.~P.}\ \bibnamefont
  {Pekola}}\ and\ \bibinfo {author} {\bibfnamefont {B.}~\bibnamefont
  {Karimi}},\ }\bibfield  {title} {\bibinfo {title} {Colloquium: Quantum heat
  transport in condensed matter systems},\ }\href
  {https://doi.org/10.1103/RevModPhys.93.041001} {\bibfield  {journal}
  {\bibinfo  {journal} {Rev. Mod. Phys.}\ }\textbf {\bibinfo {volume} {93}},\
  \bibinfo {pages} {041001} (\bibinfo {year} {2021})}\BibitemShut {NoStop}%
\bibitem [{\citenamefont {Bauer}\ \emph {et~al.}(2012)\citenamefont {Bauer},
  \citenamefont {Saitoh},\ and\ \citenamefont {Wees}}]{bauer_spin_2012}%
  \BibitemOpen
  \bibfield  {author} {\bibinfo {author} {\bibfnamefont {G.~E.~W.}\
  \bibnamefont {Bauer}}, \bibinfo {author} {\bibfnamefont {E.}~\bibnamefont
  {Saitoh}},\ and\ \bibinfo {author} {\bibfnamefont {B.~J.~v.}\ \bibnamefont
  {Wees}},\ }\bibfield  {title} {\bibinfo {title} {Spin caloritronics},\ }\href
  {https://doi.org/10.1038/nmat3301} {\bibfield  {journal} {\bibinfo  {journal}
  {Nat. Mater.}\ }\textbf {\bibinfo {volume} {11}},\ \bibinfo {pages} {391}
  (\bibinfo {year} {2012})}\BibitemShut {NoStop}%
\bibitem [{\citenamefont {Fornieri}\ and\ \citenamefont
  {Giazotto}(2017)}]{fornieri_towards_2017}%
  \BibitemOpen
  \bibfield  {author} {\bibinfo {author} {\bibfnamefont {A.}~\bibnamefont
  {Fornieri}}\ and\ \bibinfo {author} {\bibfnamefont {F.}~\bibnamefont
  {Giazotto}},\ }\bibfield  {title} {\bibinfo {title} {Towards phase-coherent
  caloritronics in superconducting circuits},\ }\href
  {https://doi.org/10.1038/nnano.2017.204} {\bibfield  {journal} {\bibinfo
  {journal} {Nature Nanotech.}\ }\textbf {\bibinfo {volume} {12}},\ \bibinfo
  {pages} {944} (\bibinfo {year} {2017})}\BibitemShut {NoStop}%
\bibitem [{\citenamefont {Hwang}\ and\ \citenamefont
  {Sothmann}(2020)}]{Hwang2020}%
  \BibitemOpen
  \bibfield  {author} {\bibinfo {author} {\bibfnamefont {S.-Y.}\ \bibnamefont
  {Hwang}}\ and\ \bibinfo {author} {\bibfnamefont {B.}~\bibnamefont
  {Sothmann}},\ }\bibfield  {title} {\bibinfo {title} {Phase-coherent
  caloritronics with ordinary and topological josephson junctions},\ }\href
  {https://doi.org/10.1140/epjst/e2019-900094-y} {\bibfield  {journal}
  {\bibinfo  {journal} {The European Physical Journal Special Topics}\ }\textbf
  {\bibinfo {volume} {229}},\ \bibinfo {pages} {683} (\bibinfo {year}
  {2020})}\BibitemShut {NoStop}%
\bibitem [{\citenamefont {Sothmann}\ \emph {et~al.}(2015)\citenamefont
  {Sothmann}, \citenamefont {S\'anchez},\ and\ \citenamefont
  {Jordan}}]{sothmann_thermoelectric_2015}%
  \BibitemOpen
  \bibfield  {author} {\bibinfo {author} {\bibfnamefont {B.}~\bibnamefont
  {Sothmann}}, \bibinfo {author} {\bibfnamefont {R.}~\bibnamefont
  {S\'anchez}},\ and\ \bibinfo {author} {\bibfnamefont {A.~N.}\ \bibnamefont
  {Jordan}},\ }\bibfield  {title} {\bibinfo {title} {Thermoelectric energy
  harvesting with quantum dots},\ }\href
  {https://doi.org/10.1088/0957-4484/26/3/032001} {\bibfield  {journal}
  {\bibinfo  {journal} {Nanotechnology}\ }\textbf {\bibinfo {volume} {26}},\
  \bibinfo {pages} {032001} (\bibinfo {year} {2015})}\BibitemShut {NoStop}%
\bibitem [{\citenamefont {Benenti}\ \emph {et~al.}(2017)\citenamefont
  {Benenti}, \citenamefont {Casati}, \citenamefont {Saito},\ and\ \citenamefont
  {Whitney}}]{benenti2017fundamental}%
  \BibitemOpen
  \bibfield  {author} {\bibinfo {author} {\bibfnamefont {G.}~\bibnamefont
  {Benenti}}, \bibinfo {author} {\bibfnamefont {G.}~\bibnamefont {Casati}},
  \bibinfo {author} {\bibfnamefont {K.}~\bibnamefont {Saito}},\ and\ \bibinfo
  {author} {\bibfnamefont {R.~S.}\ \bibnamefont {Whitney}},\ }\bibfield
  {title} {\bibinfo {title} {Fundamental aspects of steady-state conversion of
  heat to work at the nanoscale},\ }\href@noop {} {\bibfield  {journal}
  {\bibinfo  {journal} {Physics Reports}\ }\textbf {\bibinfo {volume} {694}},\
  \bibinfo {pages} {1} (\bibinfo {year} {2017})}\BibitemShut {NoStop}%
\bibitem [{\citenamefont {Arrachea}(2023)}]{arrachea2022energy}%
  \BibitemOpen
  \bibfield  {author} {\bibinfo {author} {\bibfnamefont {L.~d.~C.}\
  \bibnamefont {Arrachea}},\ }\bibfield  {title} {\bibinfo {title} {Energy
  dynamics, heat production and heat--work conversion with qubits: towards the
  development of quantum machines},\ }\href@noop {} {\bibfield  {journal}
  {\bibinfo  {journal} {Reports on Progress in Physics}\ } (\bibinfo {year}
  {2023})}\BibitemShut {NoStop}%
\bibitem [{\citenamefont {Giazotto}\ and\ \citenamefont
  {Martínez-Pérez}(2012)}]{giazotto_phase-controlled_2012}%
  \BibitemOpen
  \bibfield  {author} {\bibinfo {author} {\bibfnamefont {F.}~\bibnamefont
  {Giazotto}}\ and\ \bibinfo {author} {\bibfnamefont {M.~J.}\ \bibnamefont
  {Martínez-Pérez}},\ }\bibfield  {title} {\bibinfo {title} {Phase-controlled
  superconducting heat-flux quantum modulator},\ }\href
  {https://doi.org/10.1063/1.4750068} {\bibfield  {journal} {\bibinfo
  {journal} {Applied Physics Letters}\ }\textbf {\bibinfo {volume} {101}},\
  \bibinfo {pages} {102601} (\bibinfo {year} {2012})}\BibitemShut {NoStop}%
\bibitem [{\citenamefont {Giazotto}\ and\ \citenamefont
  {Mart\'inez-P\'erez}(2012)}]{giazotto_josephson_2012}%
  \BibitemOpen
  \bibfield  {author} {\bibinfo {author} {\bibfnamefont {F.}~\bibnamefont
  {Giazotto}}\ and\ \bibinfo {author} {\bibfnamefont {M.~J.}\ \bibnamefont
  {Mart\'inez-P\'erez}},\ }\bibfield  {title} {\bibinfo {title} {The
  {Josephson} heat interferometer},\ }\href
  {https://doi.org/10.1038/nature11702} {\bibfield  {journal} {\bibinfo
  {journal} {Nature}\ }\textbf {\bibinfo {volume} {492}},\ \bibinfo {pages}
  {401} (\bibinfo {year} {2012})}\BibitemShut {NoStop}%
\bibitem [{\citenamefont {Martínez-Pérez}\ and\ \citenamefont
  {Giazotto}(2013{\natexlab{a}})}]{martinez-perez_fully_2013}%
  \BibitemOpen
  \bibfield  {author} {\bibinfo {author} {\bibfnamefont {M.~J.}\ \bibnamefont
  {Martínez-Pérez}}\ and\ \bibinfo {author} {\bibfnamefont {F.}~\bibnamefont
  {Giazotto}},\ }\bibfield  {title} {\bibinfo {title} {Fully balanced heat
  interferometer},\ }\href {https://doi.org/doi:10.1063/1.4794412} {\bibfield
  {journal} {\bibinfo  {journal} {Applied Physics Letters}\ }\textbf {\bibinfo
  {volume} {102}},\ \bibinfo {pages} {092602} (\bibinfo {year}
  {2013}{\natexlab{a}})}\BibitemShut {NoStop}%
\bibitem [{\citenamefont {Fornieri}\ \emph
  {et~al.}(2016{\natexlab{a}})\citenamefont {Fornieri}, \citenamefont {Blanc},
  \citenamefont {Bosisio}, \citenamefont {D'Ambrosio},\ and\ \citenamefont
  {Giazotto}}]{fornieri_nanoscale_2016}%
  \BibitemOpen
  \bibfield  {author} {\bibinfo {author} {\bibfnamefont {A.}~\bibnamefont
  {Fornieri}}, \bibinfo {author} {\bibfnamefont {C.}~\bibnamefont {Blanc}},
  \bibinfo {author} {\bibfnamefont {R.}~\bibnamefont {Bosisio}}, \bibinfo
  {author} {\bibfnamefont {S.}~\bibnamefont {D'Ambrosio}},\ and\ \bibinfo
  {author} {\bibfnamefont {F.}~\bibnamefont {Giazotto}},\ }\bibfield  {title}
  {\bibinfo {title} {Nanoscale phase engineering of thermal transport with a
  {Josephson} heat modulator},\ }\href {https://doi.org/10.1038/nnano.2015.281}
  {\bibfield  {journal} {\bibinfo  {journal} {Nat Nano}\ }\textbf {\bibinfo
  {volume} {11}},\ \bibinfo {pages} {258} (\bibinfo {year}
  {2016}{\natexlab{a}})}\BibitemShut {NoStop}%
\bibitem [{\citenamefont {Fornieri}\ \emph {et~al.}(2017)\citenamefont
  {Fornieri}, \citenamefont {Timossi}, \citenamefont {Virtanen}, \citenamefont
  {Solinas},\ and\ \citenamefont {Giazotto}}]{fornieri_0-pi_2017}%
  \BibitemOpen
  \bibfield  {author} {\bibinfo {author} {\bibfnamefont {A.}~\bibnamefont
  {Fornieri}}, \bibinfo {author} {\bibfnamefont {G.}~\bibnamefont {Timossi}},
  \bibinfo {author} {\bibfnamefont {P.}~\bibnamefont {Virtanen}}, \bibinfo
  {author} {\bibfnamefont {P.}~\bibnamefont {Solinas}},\ and\ \bibinfo {author}
  {\bibfnamefont {F.}~\bibnamefont {Giazotto}},\ }\bibfield  {title} {\bibinfo
  {title} {0-pi phase-controllable thermal {Josephson} junction},\ }\href
  {https://doi.org/10.1038/nnano.2017.25} {\bibfield  {journal} {\bibinfo
  {journal} {Nature Nanotechnology}\ }\textbf {\bibinfo {volume} {12}},\
  \bibinfo {pages} {425} (\bibinfo {year} {2017})}\BibitemShut {NoStop}%
\bibitem [{\citenamefont {Martínez-Pérez}\ and\ \citenamefont
  {Giazotto}(2013{\natexlab{b}})}]{martinez-perez_efficient_2013}%
  \BibitemOpen
  \bibfield  {author} {\bibinfo {author} {\bibfnamefont {M.~J.}\ \bibnamefont
  {Martínez-Pérez}}\ and\ \bibinfo {author} {\bibfnamefont {F.}~\bibnamefont
  {Giazotto}},\ }\bibfield  {title} {\bibinfo {title} {Efficient phase-tunable
  {Josephson} thermal rectifier},\ }\href
  {https://doi.org/doi:10.1063/1.4804550} {\bibfield  {journal} {\bibinfo
  {journal} {Applied Physics Letters}\ }\textbf {\bibinfo {volume} {102}},\
  \bibinfo {pages} {182602} (\bibinfo {year} {2013}{\natexlab{b}})}\BibitemShut
  {NoStop}%
\bibitem [{\citenamefont {Giazotto}\ and\ \citenamefont
  {Bergeret}(2013)}]{giazotto_thermal_2013}%
  \BibitemOpen
  \bibfield  {author} {\bibinfo {author} {\bibfnamefont {F.}~\bibnamefont
  {Giazotto}}\ and\ \bibinfo {author} {\bibfnamefont {F.~S.}\ \bibnamefont
  {Bergeret}},\ }\bibfield  {title} {\bibinfo {title} {Thermal rectification of
  electrons in hybrid normal metal-superconductor nanojunctions},\ }\href
  {https://doi.org/10.1063/1.4846375} {\bibfield  {journal} {\bibinfo
  {journal} {Applied Physics Letters}\ }\textbf {\bibinfo {volume} {103}},\
  \bibinfo {pages} {242602} (\bibinfo {year} {2013})}\BibitemShut {NoStop}%
\bibitem [{\citenamefont {Fornieri}\ \emph {et~al.}(2014)\citenamefont
  {Fornieri}, \citenamefont {Martínez-Pérez},\ and\ \citenamefont
  {Giazotto}}]{fornieri_normal_2014}%
  \BibitemOpen
  \bibfield  {author} {\bibinfo {author} {\bibfnamefont {A.}~\bibnamefont
  {Fornieri}}, \bibinfo {author} {\bibfnamefont {M.~J.}\ \bibnamefont
  {Martínez-Pérez}},\ and\ \bibinfo {author} {\bibfnamefont {F.}~\bibnamefont
  {Giazotto}},\ }\bibfield  {title} {\bibinfo {title} {A normal metal
  tunnel-junction heat diode},\ }\href {https://doi.org/10.1063/1.4875917}
  {\bibfield  {journal} {\bibinfo  {journal} {Applied Physics Letters}\
  }\textbf {\bibinfo {volume} {104}},\ \bibinfo {pages} {183108} (\bibinfo
  {year} {2014})}\BibitemShut {NoStop}%
\bibitem [{\citenamefont {Martínez-Pérez}\ \emph {et~al.}(2015)\citenamefont
  {Martínez-Pérez}, \citenamefont {Fornieri},\ and\ \citenamefont
  {Giazotto}}]{martinez-perez_rectification_2015}%
  \BibitemOpen
  \bibfield  {author} {\bibinfo {author} {\bibfnamefont {M.~J.}\ \bibnamefont
  {Martínez-Pérez}}, \bibinfo {author} {\bibfnamefont {A.}~\bibnamefont
  {Fornieri}},\ and\ \bibinfo {author} {\bibfnamefont {F.}~\bibnamefont
  {Giazotto}},\ }\bibfield  {title} {\bibinfo {title} {Rectification of
  electronic heat current by a hybrid thermal diode},\ }\href
  {https://doi.org/10.1038/nnano.2015.11} {\bibfield  {journal} {\bibinfo
  {journal} {Nat Nano}\ }\textbf {\bibinfo {volume} {10}},\ \bibinfo {pages}
  {303} (\bibinfo {year} {2015})}\BibitemShut {NoStop}%
\bibitem [{\citenamefont {Giazotto}\ \emph {et~al.}(2014)\citenamefont
  {Giazotto}, \citenamefont {Robinson}, \citenamefont {Moodera},\ and\
  \citenamefont {Bergeret}}]{giazotto_proposal_2014}%
  \BibitemOpen
  \bibfield  {author} {\bibinfo {author} {\bibfnamefont {F.}~\bibnamefont
  {Giazotto}}, \bibinfo {author} {\bibfnamefont {J.~W.~A.}\ \bibnamefont
  {Robinson}}, \bibinfo {author} {\bibfnamefont {J.~S.}\ \bibnamefont
  {Moodera}},\ and\ \bibinfo {author} {\bibfnamefont {F.~S.}\ \bibnamefont
  {Bergeret}},\ }\bibfield  {title} {\bibinfo {title} {Proposal for a
  phase-coherent thermoelectric transistor},\ }\href
  {https://doi.org/10.1063/1.4893443} {\bibfield  {journal} {\bibinfo
  {journal} {Appl. Phys. Lett.}\ }\textbf {\bibinfo {volume} {105}},\ \bibinfo
  {pages} {062602} (\bibinfo {year} {2014})}\BibitemShut {NoStop}%
\bibitem [{\citenamefont {Fornieri}\ \emph
  {et~al.}(2016{\natexlab{b}})\citenamefont {Fornieri}, \citenamefont
  {Timossi}, \citenamefont {Bosisio}, \citenamefont {Solinas},\ and\
  \citenamefont {Giazotto}}]{fornieri_negative_2016}%
  \BibitemOpen
  \bibfield  {author} {\bibinfo {author} {\bibfnamefont {A.}~\bibnamefont
  {Fornieri}}, \bibinfo {author} {\bibfnamefont {G.}~\bibnamefont {Timossi}},
  \bibinfo {author} {\bibfnamefont {R.}~\bibnamefont {Bosisio}}, \bibinfo
  {author} {\bibfnamefont {P.}~\bibnamefont {Solinas}},\ and\ \bibinfo {author}
  {\bibfnamefont {F.}~\bibnamefont {Giazotto}},\ }\bibfield  {title} {\bibinfo
  {title} {Negative differential thermal conductance and heat amplification in
  superconducting hybrid devices},\ }\href
  {https://doi.org/10.1103/PhysRevB.93.134508} {\bibfield  {journal} {\bibinfo
  {journal} {Phys. Rev. B}\ }\textbf {\bibinfo {volume} {93}},\ \bibinfo
  {pages} {134508} (\bibinfo {year} {2016}{\natexlab{b}})}\BibitemShut
  {NoStop}%
\bibitem [{\citenamefont {Sothmann}\ \emph {et~al.}(2017)\citenamefont
  {Sothmann}, \citenamefont {Giazotto},\ and\ \citenamefont
  {Hankiewicz}}]{sothmann_high-efficiency_2017}%
  \BibitemOpen
  \bibfield  {author} {\bibinfo {author} {\bibfnamefont {B.}~\bibnamefont
  {Sothmann}}, \bibinfo {author} {\bibfnamefont {F.}~\bibnamefont {Giazotto}},\
  and\ \bibinfo {author} {\bibfnamefont {E.~M.}\ \bibnamefont {Hankiewicz}},\
  }\bibfield  {title} {\bibinfo {title} {High-efficiency thermal switch based
  on topological {Josephson} junctions},\ }\href
  {https://doi.org/10.1088/1367-2630/aa60d4} {\bibfield  {journal} {\bibinfo
  {journal} {New J. Phys.}\ }\textbf {\bibinfo {volume} {19}},\ \bibinfo
  {pages} {023056} (\bibinfo {year} {2017})}\BibitemShut {NoStop}%
\bibitem [{\citenamefont {Hwang}\ \emph {et~al.}(2018)\citenamefont {Hwang},
  \citenamefont {Giazotto},\ and\ \citenamefont
  {Sothmann}}]{hwang_phase-coherent_2018}%
  \BibitemOpen
  \bibfield  {author} {\bibinfo {author} {\bibfnamefont {S.-Y.}\ \bibnamefont
  {Hwang}}, \bibinfo {author} {\bibfnamefont {F.}~\bibnamefont {Giazotto}},\
  and\ \bibinfo {author} {\bibfnamefont {B.}~\bibnamefont {Sothmann}},\
  }\bibfield  {title} {\bibinfo {title} {Phase-{Coherent} {Heat} {Circulator}
  {Based} on {Multiterminal} {Josephson} {Junctions}},\ }\href
  {https://doi.org/10.1103/PhysRevApplied.10.044062} {\bibfield  {journal}
  {\bibinfo  {journal} {Phys. Rev. Applied}\ }\textbf {\bibinfo {volume}
  {10}},\ \bibinfo {pages} {044062} (\bibinfo {year} {2018})}\BibitemShut
  {NoStop}%
\bibitem [{\citenamefont {Acciai}\ \emph {et~al.}(2021)\citenamefont {Acciai},
  \citenamefont {Hajiloo}, \citenamefont {Hassler},\ and\ \citenamefont
  {Splettstoesser}}]{acciai_phase-coherent_2021}%
  \BibitemOpen
  \bibfield  {author} {\bibinfo {author} {\bibfnamefont {M.}~\bibnamefont
  {Acciai}}, \bibinfo {author} {\bibfnamefont {F.}~\bibnamefont {Hajiloo}},
  \bibinfo {author} {\bibfnamefont {F.}~\bibnamefont {Hassler}},\ and\ \bibinfo
  {author} {\bibfnamefont {J.}~\bibnamefont {Splettstoesser}},\ }\bibfield
  {title} {\bibinfo {title} {Phase-coherent heat circulators with normal or
  superconducting contacts},\ }\href
  {https://doi.org/10.1103/PhysRevB.103.085409} {\bibfield  {journal} {\bibinfo
   {journal} {Phys. Rev. B}\ }\textbf {\bibinfo {volume} {103}},\ \bibinfo
  {pages} {085409} (\bibinfo {year} {2021})},\ \bibinfo {note} {publisher:
  American Physical Society}\BibitemShut {NoStop}%
\bibitem [{\citenamefont {Guarcello}\ \emph {et~al.}(2018)\citenamefont
  {Guarcello}, \citenamefont {Solinas}, \citenamefont {Braggio}, \citenamefont
  {Di~Ventra},\ and\ \citenamefont {Giazotto}}]{guarcello_josephson_2018}%
  \BibitemOpen
  \bibfield  {author} {\bibinfo {author} {\bibfnamefont {C.}~\bibnamefont
  {Guarcello}}, \bibinfo {author} {\bibfnamefont {P.}~\bibnamefont {Solinas}},
  \bibinfo {author} {\bibfnamefont {A.}~\bibnamefont {Braggio}}, \bibinfo
  {author} {\bibfnamefont {M.}~\bibnamefont {Di~Ventra}},\ and\ \bibinfo
  {author} {\bibfnamefont {F.}~\bibnamefont {Giazotto}},\ }\bibfield  {title}
  {\bibinfo {title} {Josephson {Thermal} {Memory}},\ }\href
  {https://doi.org/10.1103/PhysRevApplied.9.014021} {\bibfield  {journal}
  {\bibinfo  {journal} {Phys. Rev. Applied}\ }\textbf {\bibinfo {volume} {9}},\
  \bibinfo {pages} {014021} (\bibinfo {year} {2018})}\BibitemShut {NoStop}%
\bibitem [{\citenamefont {Solinas}\ \emph {et~al.}(2016)\citenamefont
  {Solinas}, \citenamefont {Bosisio},\ and\ \citenamefont
  {Giazotto}}]{solinas_microwave_2016}%
  \BibitemOpen
  \bibfield  {author} {\bibinfo {author} {\bibfnamefont {P.}~\bibnamefont
  {Solinas}}, \bibinfo {author} {\bibfnamefont {R.}~\bibnamefont {Bosisio}},\
  and\ \bibinfo {author} {\bibfnamefont {F.}~\bibnamefont {Giazotto}},\
  }\bibfield  {title} {\bibinfo {title} {Microwave quantum refrigeration based
  on the {Josephson} effect},\ }\href
  {https://doi.org/10.1103/PhysRevB.93.224521} {\bibfield  {journal} {\bibinfo
  {journal} {Phys. Rev. B}\ }\textbf {\bibinfo {volume} {93}},\ \bibinfo
  {pages} {224521} (\bibinfo {year} {2016})}\BibitemShut {NoStop}%
\bibitem [{\citenamefont {Hofer}\ \emph {et~al.}(2016)\citenamefont {Hofer},
  \citenamefont {Perarnau-Llobet}, \citenamefont {Brask}, \citenamefont
  {Silva}, \citenamefont {Huber},\ and\ \citenamefont
  {Brunner}}]{hofer_autonomous_2016}%
  \BibitemOpen
  \bibfield  {author} {\bibinfo {author} {\bibfnamefont {P.~P.}\ \bibnamefont
  {Hofer}}, \bibinfo {author} {\bibfnamefont {M.}~\bibnamefont
  {Perarnau-Llobet}}, \bibinfo {author} {\bibfnamefont {J.~B.}\ \bibnamefont
  {Brask}}, \bibinfo {author} {\bibfnamefont {R.}~\bibnamefont {Silva}},
  \bibinfo {author} {\bibfnamefont {M.}~\bibnamefont {Huber}},\ and\ \bibinfo
  {author} {\bibfnamefont {N.}~\bibnamefont {Brunner}},\ }\bibfield  {title}
  {\bibinfo {title} {Autonomous quantum refrigerator in a circuit {QED}
  architecture based on a {Josephson} junction},\ }\href
  {https://doi.org/10.1103/PhysRevB.94.235420} {\bibfield  {journal} {\bibinfo
  {journal} {Phys. Rev. B}\ }\textbf {\bibinfo {volume} {94}},\ \bibinfo
  {pages} {235420} (\bibinfo {year} {2016})}\BibitemShut {NoStop}%
\bibitem [{\citenamefont {Vischi}\ \emph {et~al.}(2017)\citenamefont {Vischi},
  \citenamefont {Carrega}, \citenamefont {Strambini}, \citenamefont
  {D'Ambrosio}, \citenamefont {Bergeret}, \citenamefont {Nazarov},\ and\
  \citenamefont {Giazotto}}]{vischi_coherent_2017}%
  \BibitemOpen
  \bibfield  {author} {\bibinfo {author} {\bibfnamefont {F.}~\bibnamefont
  {Vischi}}, \bibinfo {author} {\bibfnamefont {M.}~\bibnamefont {Carrega}},
  \bibinfo {author} {\bibfnamefont {E.}~\bibnamefont {Strambini}}, \bibinfo
  {author} {\bibfnamefont {S.}~\bibnamefont {D'Ambrosio}}, \bibinfo {author}
  {\bibfnamefont {F.~S.}\ \bibnamefont {Bergeret}}, \bibinfo {author}
  {\bibfnamefont {Y.~V.}\ \bibnamefont {Nazarov}},\ and\ \bibinfo {author}
  {\bibfnamefont {F.}~\bibnamefont {Giazotto}},\ }\bibfield  {title} {\bibinfo
  {title} {Coherent transport properties of a three-terminal hybrid
  superconducting interferometer},\ }\href
  {https://doi.org/10.1103/PhysRevB.95.054504} {\bibfield  {journal} {\bibinfo
  {journal} {Phys. Rev. B}\ }\textbf {\bibinfo {volume} {95}},\ \bibinfo
  {pages} {054504} (\bibinfo {year} {2017})}\BibitemShut {NoStop}%
\bibitem [{\citenamefont {Edwards}\ \emph {et~al.}(1993)\citenamefont
  {Edwards}, \citenamefont {Niu},\ and\ \citenamefont
  {de~Lozanne}}]{edwards_a_1993}%
  \BibitemOpen
  \bibfield  {author} {\bibinfo {author} {\bibfnamefont {H.~L.}\ \bibnamefont
  {Edwards}}, \bibinfo {author} {\bibfnamefont {Q.}~\bibnamefont {Niu}},\ and\
  \bibinfo {author} {\bibfnamefont {A.~L.}\ \bibnamefont {de~Lozanne}},\
  }\bibfield  {title} {\bibinfo {title} {A quantum‐dot refrigerator},\ }\href
  {https://doi.org/10.1063/1.110672} {\bibfield  {journal} {\bibinfo  {journal}
  {Applied Physics Letters}\ }\textbf {\bibinfo {volume} {63}},\ \bibinfo
  {pages} {1815} (\bibinfo {year} {1993})},\ \Eprint
  {https://arxiv.org/abs/https://doi.org/10.1063/1.110672}
  {https://doi.org/10.1063/1.110672} \BibitemShut {NoStop}%
\bibitem [{\citenamefont {Edwards}\ \emph {et~al.}(1995)\citenamefont
  {Edwards}, \citenamefont {Niu}, \citenamefont {Georgakis},\ and\
  \citenamefont {de~Lozanne}}]{edwards_cryogenic_1995}%
  \BibitemOpen
  \bibfield  {author} {\bibinfo {author} {\bibfnamefont {H.~L.}\ \bibnamefont
  {Edwards}}, \bibinfo {author} {\bibfnamefont {Q.}~\bibnamefont {Niu}},
  \bibinfo {author} {\bibfnamefont {G.~A.}\ \bibnamefont {Georgakis}},\ and\
  \bibinfo {author} {\bibfnamefont {A.~L.}\ \bibnamefont {de~Lozanne}},\
  }\bibfield  {title} {\bibinfo {title} {Cryogenic cooling using tunneling
  structures with sharp energy features},\ }\href
  {https://doi.org/10.1103/PhysRevB.52.5714} {\bibfield  {journal} {\bibinfo
  {journal} {Phys. Rev. B}\ }\textbf {\bibinfo {volume} {52}},\ \bibinfo
  {pages} {5714} (\bibinfo {year} {1995})}\BibitemShut {NoStop}%
\bibitem [{\citenamefont {Prance}\ \emph {et~al.}(2009)\citenamefont {Prance},
  \citenamefont {Smith}, \citenamefont {Griffiths}, \citenamefont {Chorley},
  \citenamefont {Anderson}, \citenamefont {Jones}, \citenamefont {Farrer},\
  and\ \citenamefont {Ritchie}}]{prance_electronic_2009}%
  \BibitemOpen
  \bibfield  {author} {\bibinfo {author} {\bibfnamefont {J.~R.}\ \bibnamefont
  {Prance}}, \bibinfo {author} {\bibfnamefont {C.~G.}\ \bibnamefont {Smith}},
  \bibinfo {author} {\bibfnamefont {J.~P.}\ \bibnamefont {Griffiths}}, \bibinfo
  {author} {\bibfnamefont {S.~J.}\ \bibnamefont {Chorley}}, \bibinfo {author}
  {\bibfnamefont {D.}~\bibnamefont {Anderson}}, \bibinfo {author}
  {\bibfnamefont {G.~A.~C.}\ \bibnamefont {Jones}}, \bibinfo {author}
  {\bibfnamefont {I.}~\bibnamefont {Farrer}},\ and\ \bibinfo {author}
  {\bibfnamefont {D.~A.}\ \bibnamefont {Ritchie}},\ }\bibfield  {title}
  {\bibinfo {title} {Electronic {Refrigeration} of a {Two}-{Dimensional}
  {Electron} {Gas}},\ }\href {https://doi.org/10.1103/PhysRevLett.102.146602}
  {\bibfield  {journal} {\bibinfo  {journal} {Phys. Rev. Lett.}\ }\textbf
  {\bibinfo {volume} {102}},\ \bibinfo {pages} {146602} (\bibinfo {year}
  {2009})}\BibitemShut {NoStop}%
\bibitem [{\citenamefont {Gasparinetti}\ \emph {et~al.}(2011)\citenamefont
  {Gasparinetti}, \citenamefont {Deon}, \citenamefont {Biasiol}, \citenamefont
  {Sorba}, \citenamefont {Beltram},\ and\ \citenamefont
  {Giazotto}}]{gasparinetti_probing_2011}%
  \BibitemOpen
  \bibfield  {author} {\bibinfo {author} {\bibfnamefont {S.}~\bibnamefont
  {Gasparinetti}}, \bibinfo {author} {\bibfnamefont {F.}~\bibnamefont {Deon}},
  \bibinfo {author} {\bibfnamefont {G.}~\bibnamefont {Biasiol}}, \bibinfo
  {author} {\bibfnamefont {L.}~\bibnamefont {Sorba}}, \bibinfo {author}
  {\bibfnamefont {F.}~\bibnamefont {Beltram}},\ and\ \bibinfo {author}
  {\bibfnamefont {F.}~\bibnamefont {Giazotto}},\ }\bibfield  {title} {\bibinfo
  {title} {Probing the local temperature of a two-dimensional electron gas
  microdomain with a quantum dot: {Measurement} of electron-phonon
  interaction},\ }\href {https://doi.org/10.1103/PhysRevB.83.201306} {\bibfield
   {journal} {\bibinfo  {journal} {Phys. Rev. B}\ }\textbf {\bibinfo {volume}
  {83}},\ \bibinfo {pages} {201306} (\bibinfo {year} {2011})}\BibitemShut
  {NoStop}%
\bibitem [{\citenamefont {Timofeev}\ \emph {et~al.}(2009)\citenamefont
  {Timofeev}, \citenamefont {Helle}, \citenamefont {Meschke}, \citenamefont
  {Möttönen},\ and\ \citenamefont {Pekola}}]{timofeev_electronic_2009}%
  \BibitemOpen
  \bibfield  {author} {\bibinfo {author} {\bibfnamefont {A.~V.}\ \bibnamefont
  {Timofeev}}, \bibinfo {author} {\bibfnamefont {M.}~\bibnamefont {Helle}},
  \bibinfo {author} {\bibfnamefont {M.}~\bibnamefont {Meschke}}, \bibinfo
  {author} {\bibfnamefont {M.}~\bibnamefont {Möttönen}},\ and\ \bibinfo
  {author} {\bibfnamefont {J.~P.}\ \bibnamefont {Pekola}},\ }\bibfield  {title}
  {\bibinfo {title} {Electronic {Refrigeration} at the {Quantum} {Limit}},\
  }\href {https://doi.org/10.1103/PhysRevLett.102.200801} {\bibfield  {journal}
  {\bibinfo  {journal} {Phys. Rev. Lett.}\ }\textbf {\bibinfo {volume} {102}},\
  \bibinfo {pages} {200801} (\bibinfo {year} {2009})}\BibitemShut {NoStop}%
\bibitem [{\citenamefont {Arrachea}\ \emph {et~al.}(2007)\citenamefont
  {Arrachea}, \citenamefont {Moskalets},\ and\ \citenamefont
  {Martin-Moreno}}]{arrachea_heat_2007}%
  \BibitemOpen
  \bibfield  {author} {\bibinfo {author} {\bibfnamefont {L.}~\bibnamefont
  {Arrachea}}, \bibinfo {author} {\bibfnamefont {M.}~\bibnamefont
  {Moskalets}},\ and\ \bibinfo {author} {\bibfnamefont {L.}~\bibnamefont
  {Martin-Moreno}},\ }\bibfield  {title} {\bibinfo {title} {Heat production and
  energy balance in nanoscale engines driven by time-dependent fields},\ }\href
  {https://doi.org/10.1103/PhysRevB.75.245420} {\bibfield  {journal} {\bibinfo
  {journal} {Phys. Rev. B}\ }\textbf {\bibinfo {volume} {75}},\ \bibinfo
  {pages} {245420} (\bibinfo {year} {2007})}\BibitemShut {NoStop}%
\bibitem [{\citenamefont {Rey}\ \emph {et~al.}(2007)\citenamefont {Rey},
  \citenamefont {Strass}, \citenamefont {Kohler}, \citenamefont {Hänggi},\
  and\ \citenamefont {Sols}}]{rey_nonadiabatic_2007}%
  \BibitemOpen
  \bibfield  {author} {\bibinfo {author} {\bibfnamefont {M.}~\bibnamefont
  {Rey}}, \bibinfo {author} {\bibfnamefont {M.}~\bibnamefont {Strass}},
  \bibinfo {author} {\bibfnamefont {S.}~\bibnamefont {Kohler}}, \bibinfo
  {author} {\bibfnamefont {P.}~\bibnamefont {Hänggi}},\ and\ \bibinfo {author}
  {\bibfnamefont {F.}~\bibnamefont {Sols}},\ }\bibfield  {title} {\bibinfo
  {title} {Nonadiabatic electron heat pump},\ }\href
  {https://doi.org/10.1103/PhysRevB.76.085337} {\bibfield  {journal} {\bibinfo
  {journal} {Phys. Rev. B}\ }\textbf {\bibinfo {volume} {76}},\ \bibinfo
  {pages} {085337} (\bibinfo {year} {2007})}\BibitemShut {NoStop}%
\bibitem [{\citenamefont {Cleuren}\ \emph {et~al.}(2012)\citenamefont
  {Cleuren}, \citenamefont {Rutten},\ and\ \citenamefont {Van~den
  Broeck}}]{cleuren_cooling_2012}%
  \BibitemOpen
  \bibfield  {author} {\bibinfo {author} {\bibfnamefont {B.}~\bibnamefont
  {Cleuren}}, \bibinfo {author} {\bibfnamefont {B.}~\bibnamefont {Rutten}},\
  and\ \bibinfo {author} {\bibfnamefont {C.}~\bibnamefont {Van~den Broeck}},\
  }\bibfield  {title} {\bibinfo {title} {Cooling by {Heating}: {Refrigeration}
  {Powered} by {Photons}},\ }\href
  {https://doi.org/10.1103/PhysRevLett.108.120603} {\bibfield  {journal}
  {\bibinfo  {journal} {Phys. Rev. Lett.}\ }\textbf {\bibinfo {volume} {108}},\
  \bibinfo {pages} {120603} (\bibinfo {year} {2012})}\BibitemShut {NoStop}%
\bibitem [{\citenamefont {Brüggemann}\ \emph {et~al.}(2014)\citenamefont
  {Brüggemann}, \citenamefont {Weiss}, \citenamefont {Nalbach},\ and\
  \citenamefont {Thorwart}}]{bruggemann_cooling_2014}%
  \BibitemOpen
  \bibfield  {author} {\bibinfo {author} {\bibfnamefont {J.}~\bibnamefont
  {Brüggemann}}, \bibinfo {author} {\bibfnamefont {S.}~\bibnamefont {Weiss}},
  \bibinfo {author} {\bibfnamefont {P.}~\bibnamefont {Nalbach}},\ and\ \bibinfo
  {author} {\bibfnamefont {M.}~\bibnamefont {Thorwart}},\ }\bibfield  {title}
  {\bibinfo {title} {Cooling a {Magnetic} {Nanoisland} by {Spin}-{Polarized}
  {Currents}},\ }\href {https://doi.org/10.1103/PhysRevLett.113.076602}
  {\bibfield  {journal} {\bibinfo  {journal} {Phys. Rev. Lett.}\ }\textbf
  {\bibinfo {volume} {113}},\ \bibinfo {pages} {076602} (\bibinfo {year}
  {2014})}\BibitemShut {NoStop}%
\bibitem [{\citenamefont {Pekola}\ \emph {et~al.}(2014)\citenamefont {Pekola},
  \citenamefont {Koski},\ and\ \citenamefont
  {Averin}}]{pekola_refrigerator_2014}%
  \BibitemOpen
  \bibfield  {author} {\bibinfo {author} {\bibfnamefont {J.~P.}\ \bibnamefont
  {Pekola}}, \bibinfo {author} {\bibfnamefont {J.~V.}\ \bibnamefont {Koski}},\
  and\ \bibinfo {author} {\bibfnamefont {D.~V.}\ \bibnamefont {Averin}},\
  }\bibfield  {title} {\bibinfo {title} {Refrigerator based on the {Coulomb}
  barrier for single-electron tunneling},\ }\href
  {https://doi.org/10.1103/PhysRevB.89.081309} {\bibfield  {journal} {\bibinfo
  {journal} {Phys. Rev. B}\ }\textbf {\bibinfo {volume} {89}},\ \bibinfo
  {pages} {081309} (\bibinfo {year} {2014})}\BibitemShut {NoStop}%
\bibitem [{\citenamefont {Nahum}\ \emph {et~al.}(1994)\citenamefont {Nahum},
  \citenamefont {Eiles},\ and\ \citenamefont
  {Martinis}}]{nahum_electronic_1994}%
  \BibitemOpen
  \bibfield  {author} {\bibinfo {author} {\bibfnamefont {M.}~\bibnamefont
  {Nahum}}, \bibinfo {author} {\bibfnamefont {T.~M.}\ \bibnamefont {Eiles}},\
  and\ \bibinfo {author} {\bibfnamefont {J.~M.}\ \bibnamefont {Martinis}},\
  }\bibfield  {title} {\bibinfo {title} {Electronic microrefrigerator based on
  a normal‐insulator‐superconductor tunnel junction},\ }\href
  {https://doi.org/10.1063/1.112456} {\bibfield  {journal} {\bibinfo  {journal}
  {Appl. Phys. Lett.}\ }\textbf {\bibinfo {volume} {65}},\ \bibinfo {pages}
  {3123} (\bibinfo {year} {1994})}\BibitemShut {NoStop}%
\bibitem [{\citenamefont {Leivo}\ \emph {et~al.}(1996)\citenamefont {Leivo},
  \citenamefont {Pekola},\ and\ \citenamefont {Averin}}]{leivo_efficient_1996}%
  \BibitemOpen
  \bibfield  {author} {\bibinfo {author} {\bibfnamefont {M.~M.}\ \bibnamefont
  {Leivo}}, \bibinfo {author} {\bibfnamefont {J.~P.}\ \bibnamefont {Pekola}},\
  and\ \bibinfo {author} {\bibfnamefont {D.~V.}\ \bibnamefont {Averin}},\
  }\bibfield  {title} {\bibinfo {title} {Efficient {Peltier} refrigeration by a
  pair of normal metal/insulator/superconductor junctions},\ }\href
  {https://doi.org/10.1063/1.115651} {\bibfield  {journal} {\bibinfo  {journal}
  {Appl. Phys. Lett.}\ }\textbf {\bibinfo {volume} {68}},\ \bibinfo {pages}
  {1996} (\bibinfo {year} {1996})}\BibitemShut {NoStop}%
\bibitem [{\citenamefont {Muhonen}\ \emph {et~al.}(2012)\citenamefont
  {Muhonen}, \citenamefont {Meschke},\ and\ \citenamefont
  {Pekola}}]{muhonen_micrometre-scale_2012}%
  \BibitemOpen
  \bibfield  {author} {\bibinfo {author} {\bibfnamefont {J.~T.}\ \bibnamefont
  {Muhonen}}, \bibinfo {author} {\bibfnamefont {M.}~\bibnamefont {Meschke}},\
  and\ \bibinfo {author} {\bibfnamefont {J.~P.}\ \bibnamefont {Pekola}},\
  }\bibfield  {title} {\bibinfo {title} {Micrometre-scale refrigerators},\
  }\href {https://doi.org/10.1088/0034-4885/75/4/046501} {\bibfield  {journal}
  {\bibinfo  {journal} {Rep. Prog. Phys.}\ }\textbf {\bibinfo {volume} {75}},\
  \bibinfo {pages} {046501} (\bibinfo {year} {2012})}\BibitemShut {NoStop}%
\bibitem [{\citenamefont {Rouco}\ \emph {et~al.}(2018)\citenamefont {Rouco},
  \citenamefont {Heikkil\"a},\ and\ \citenamefont
  {Bergeret}}]{rouco_electron_2018}%
  \BibitemOpen
  \bibfield  {author} {\bibinfo {author} {\bibfnamefont {M.}~\bibnamefont
  {Rouco}}, \bibinfo {author} {\bibfnamefont {T.~T.}\ \bibnamefont
  {Heikkil\"a}},\ and\ \bibinfo {author} {\bibfnamefont {F.~S.}\ \bibnamefont
  {Bergeret}},\ }\bibfield  {title} {\bibinfo {title} {Electron refrigeration
  in hybrid structures with spin-split superconductors},\ }\href
  {https://doi.org/10.1103/PhysRevB.97.014529} {\bibfield  {journal} {\bibinfo
  {journal} {Phys. Rev. B}\ }\textbf {\bibinfo {volume} {97}},\ \bibinfo
  {pages} {014529} (\bibinfo {year} {2018})}\BibitemShut {NoStop}%
\bibitem [{\citenamefont {Linden}\ \emph {et~al.}(2010)\citenamefont {Linden},
  \citenamefont {Popescu},\ and\ \citenamefont {Skrzypczyk}}]{linden_how_2010}%
  \BibitemOpen
  \bibfield  {author} {\bibinfo {author} {\bibfnamefont {N.}~\bibnamefont
  {Linden}}, \bibinfo {author} {\bibfnamefont {S.}~\bibnamefont {Popescu}},\
  and\ \bibinfo {author} {\bibfnamefont {P.}~\bibnamefont {Skrzypczyk}},\
  }\bibfield  {title} {\bibinfo {title} {How {Small} {Can} {Thermal} {Machines}
  {Be}? {The} {Smallest} {Possible} {Refrigerator}},\ }\href
  {https://doi.org/10.1103/PhysRevLett.105.130401} {\bibfield  {journal}
  {\bibinfo  {journal} {Phys. Rev. Lett.}\ }\textbf {\bibinfo {volume} {105}},\
  \bibinfo {pages} {130401} (\bibinfo {year} {2010})}\BibitemShut {NoStop}%
\bibitem [{\citenamefont {Brunner}\ \emph {et~al.}(2012)\citenamefont
  {Brunner}, \citenamefont {Linden}, \citenamefont {Popescu},\ and\
  \citenamefont {Skrzypczyk}}]{brunner_virtual_2012}%
  \BibitemOpen
  \bibfield  {author} {\bibinfo {author} {\bibfnamefont {N.}~\bibnamefont
  {Brunner}}, \bibinfo {author} {\bibfnamefont {N.}~\bibnamefont {Linden}},
  \bibinfo {author} {\bibfnamefont {S.}~\bibnamefont {Popescu}},\ and\ \bibinfo
  {author} {\bibfnamefont {P.}~\bibnamefont {Skrzypczyk}},\ }\bibfield  {title}
  {\bibinfo {title} {Virtual qubits, virtual temperatures, and the foundations
  of thermodynamics},\ }\href {https://doi.org/10.1103/PhysRevE.85.051117}
  {\bibfield  {journal} {\bibinfo  {journal} {Phys. Rev. E}\ }\textbf {\bibinfo
  {volume} {85}},\ \bibinfo {pages} {051117} (\bibinfo {year}
  {2012})}\BibitemShut {NoStop}%
\bibitem [{\citenamefont {Levy}\ and\ \citenamefont
  {Kosloff}(2012)}]{levy_quantum_2012}%
  \BibitemOpen
  \bibfield  {author} {\bibinfo {author} {\bibfnamefont {A.}~\bibnamefont
  {Levy}}\ and\ \bibinfo {author} {\bibfnamefont {R.}~\bibnamefont {Kosloff}},\
  }\bibfield  {title} {\bibinfo {title} {Quantum absorption refrigerator},\
  }\href {https://doi.org/10.1103/PhysRevLett.108.070604} {\bibfield  {journal}
  {\bibinfo  {journal} {Phys. Rev. Lett.}\ }\textbf {\bibinfo {volume} {108}},\
  \bibinfo {pages} {070604} (\bibinfo {year} {2012})}\BibitemShut {NoStop}%
\bibitem [{\citenamefont {Brunner}\ \emph {et~al.}(2014)\citenamefont
  {Brunner}, \citenamefont {Huber}, \citenamefont {Linden}, \citenamefont
  {Popescu}, \citenamefont {Silva},\ and\ \citenamefont
  {Skrzypczyk}}]{brunner_entanglement_2014}%
  \BibitemOpen
  \bibfield  {author} {\bibinfo {author} {\bibfnamefont {N.}~\bibnamefont
  {Brunner}}, \bibinfo {author} {\bibfnamefont {M.}~\bibnamefont {Huber}},
  \bibinfo {author} {\bibfnamefont {N.}~\bibnamefont {Linden}}, \bibinfo
  {author} {\bibfnamefont {S.}~\bibnamefont {Popescu}}, \bibinfo {author}
  {\bibfnamefont {R.}~\bibnamefont {Silva}},\ and\ \bibinfo {author}
  {\bibfnamefont {P.}~\bibnamefont {Skrzypczyk}},\ }\bibfield  {title}
  {\bibinfo {title} {Entanglement enhances cooling in microscopic quantum
  refrigerators},\ }\href {https://doi.org/10.1103/PhysRevE.89.032115}
  {\bibfield  {journal} {\bibinfo  {journal} {Phys. Rev. E}\ }\textbf {\bibinfo
  {volume} {89}},\ \bibinfo {pages} {032115} (\bibinfo {year}
  {2014})}\BibitemShut {NoStop}%
\bibitem [{\citenamefont {Correa}\ \emph {et~al.}(2014)\citenamefont {Correa},
  \citenamefont {Palao}, \citenamefont {Alonso},\ and\ \citenamefont
  {Adesso}}]{correa_quantum-enhanced_2014}%
  \BibitemOpen
  \bibfield  {author} {\bibinfo {author} {\bibfnamefont {L.~A.}\ \bibnamefont
  {Correa}}, \bibinfo {author} {\bibfnamefont {J.~P.}\ \bibnamefont {Palao}},
  \bibinfo {author} {\bibfnamefont {D.}~\bibnamefont {Alonso}},\ and\ \bibinfo
  {author} {\bibfnamefont {G.}~\bibnamefont {Adesso}},\ }\bibfield  {title}
  {\bibinfo {title} {Quantum-enhanced absorption refrigerators},\ }\bibfield
  {journal} {\bibinfo  {journal} {Sci. Rep.}\ }\textbf {\bibinfo {volume}
  {4}},\ \href {https://doi.org/10.1038/srep03949} {10.1038/srep03949}
  (\bibinfo {year} {2014})\BibitemShut {NoStop}%
\bibitem [{\citenamefont {Correa}(2014)}]{correa_multistage_2014}%
  \BibitemOpen
  \bibfield  {author} {\bibinfo {author} {\bibfnamefont {L.~A.}\ \bibnamefont
  {Correa}},\ }\bibfield  {title} {\bibinfo {title} {Multistage quantum
  absorption heat pumps},\ }\href {https://doi.org/10.1103/PhysRevE.89.042128}
  {\bibfield  {journal} {\bibinfo  {journal} {Phys. Rev. E}\ }\textbf {\bibinfo
  {volume} {89}},\ \bibinfo {pages} {042128} (\bibinfo {year}
  {2014})}\BibitemShut {NoStop}%
\bibitem [{\citenamefont {Venturelli}\ \emph {et~al.}(2013)\citenamefont
  {Venturelli}, \citenamefont {Fazio},\ and\ \citenamefont
  {Giovannetti}}]{venturelli_minimal_2013}%
  \BibitemOpen
  \bibfield  {author} {\bibinfo {author} {\bibfnamefont {D.}~\bibnamefont
  {Venturelli}}, \bibinfo {author} {\bibfnamefont {R.}~\bibnamefont {Fazio}},\
  and\ \bibinfo {author} {\bibfnamefont {V.}~\bibnamefont {Giovannetti}},\
  }\bibfield  {title} {\bibinfo {title} {Minimal {Self}-{Contained} {Quantum}
  {Refrigeration} {Machine} {Based} on {Four} {Quantum} {Dots}},\ }\href
  {https://doi.org/10.1103/PhysRevLett.110.256801} {\bibfield  {journal}
  {\bibinfo  {journal} {Phys. Rev. Lett.}\ }\textbf {\bibinfo {volume} {110}},\
  \bibinfo {pages} {256801} (\bibinfo {year} {2013})}\BibitemShut {NoStop}%
\bibitem [{\citenamefont {Entin-Wohlman}\ \emph {et~al.}(2015)\citenamefont
  {Entin-Wohlman}, \citenamefont {Imry},\ and\ \citenamefont
  {Aharony}}]{entin-wohlman_enhanced_2015}%
  \BibitemOpen
  \bibfield  {author} {\bibinfo {author} {\bibfnamefont {O.}~\bibnamefont
  {Entin-Wohlman}}, \bibinfo {author} {\bibfnamefont {Y.}~\bibnamefont
  {Imry}},\ and\ \bibinfo {author} {\bibfnamefont {A.}~\bibnamefont
  {Aharony}},\ }\bibfield  {title} {\bibinfo {title} {Enhanced performance of
  joint cooling and energy production},\ }\href
  {https://doi.org/10.1103/PhysRevB.91.054302} {\bibfield  {journal} {\bibinfo
  {journal} {Phys. Rev. B}\ }\textbf {\bibinfo {volume} {91}},\ \bibinfo
  {pages} {054302} (\bibinfo {year} {2015})}\BibitemShut {NoStop}%
\bibitem [{\citenamefont {Sánchez}(2017)}]{sanchez_correlation-induced_2017}%
  \BibitemOpen
  \bibfield  {author} {\bibinfo {author} {\bibfnamefont {R.}~\bibnamefont
  {Sánchez}},\ }\bibfield  {title} {\bibinfo {title} {Correlation-induced
  refrigeration with superconducting single-electron transistors},\ }\href
  {https://doi.org/10.1063/1.5008481} {\bibfield  {journal} {\bibinfo
  {journal} {Appl. Phys. Lett.}\ }\textbf {\bibinfo {volume} {111}},\ \bibinfo
  {pages} {223103} (\bibinfo {year} {2017})}\BibitemShut {NoStop}%
\bibitem [{\citenamefont {Erdman}\ \emph {et~al.}(2018)\citenamefont {Erdman},
  \citenamefont {Bhandari}, \citenamefont {Fazio}, \citenamefont {Pekola},\
  and\ \citenamefont {Taddei}}]{erdman2018absorption}%
  \BibitemOpen
  \bibfield  {author} {\bibinfo {author} {\bibfnamefont {P.~A.}\ \bibnamefont
  {Erdman}}, \bibinfo {author} {\bibfnamefont {B.}~\bibnamefont {Bhandari}},
  \bibinfo {author} {\bibfnamefont {R.}~\bibnamefont {Fazio}}, \bibinfo
  {author} {\bibfnamefont {J.~P.}\ \bibnamefont {Pekola}},\ and\ \bibinfo
  {author} {\bibfnamefont {F.}~\bibnamefont {Taddei}},\ }\bibfield  {title}
  {\bibinfo {title} {Absorption refrigerators based on coulomb-coupled
  single-electron systems},\ }\href@noop {} {\bibfield  {journal} {\bibinfo
  {journal} {Physical Review B}\ }\textbf {\bibinfo {volume} {98}},\ \bibinfo
  {pages} {045433} (\bibinfo {year} {2018})}\BibitemShut {NoStop}%
\bibitem [{\citenamefont {Wang}\ \emph {et~al.}(2018)\citenamefont {Wang},
  \citenamefont {Lu}, \citenamefont {Wang},\ and\ \citenamefont
  {Jiang}}]{wang_nonlinear_2018}%
  \BibitemOpen
  \bibfield  {author} {\bibinfo {author} {\bibfnamefont {R.}~\bibnamefont
  {Wang}}, \bibinfo {author} {\bibfnamefont {J.}~\bibnamefont {Lu}}, \bibinfo
  {author} {\bibfnamefont {C.}~\bibnamefont {Wang}},\ and\ \bibinfo {author}
  {\bibfnamefont {J.-H.}\ \bibnamefont {Jiang}},\ }\bibfield  {title} {\bibinfo
  {title} {Nonlinear effects for three-terminal heat engine and refrigerator},\
  }\href {https://doi.org/10.1038/s41598-018-20757-8} {\bibfield  {journal}
  {\bibinfo  {journal} {Sci. Rep.}\ }\textbf {\bibinfo {volume} {8}},\ \bibinfo
  {pages} {2607} (\bibinfo {year} {2018})}\BibitemShut {NoStop}%
\bibitem [{\citenamefont {Sánchez}\ \emph {et~al.}(2018)\citenamefont
  {Sánchez}, \citenamefont {Burset},\ and\ \citenamefont
  {Yeyati}}]{sanchez_cooling_2018}%
  \BibitemOpen
  \bibfield  {author} {\bibinfo {author} {\bibfnamefont {R.}~\bibnamefont
  {Sánchez}}, \bibinfo {author} {\bibfnamefont {P.}~\bibnamefont {Burset}},\
  and\ \bibinfo {author} {\bibfnamefont {A.~L.}\ \bibnamefont {Yeyati}},\
  }\bibfield  {title} {\bibinfo {title} {Cooling by {Cooper} pair splitting},\
  }\href {https://doi.org/10.1103/PhysRevB.98.241414} {\bibfield  {journal}
  {\bibinfo  {journal} {Phys. Rev. B}\ }\textbf {\bibinfo {volume} {98}},\
  \bibinfo {pages} {241414} (\bibinfo {year} {2018})}\BibitemShut {NoStop}%
\bibitem [{\citenamefont {Hussein}\ \emph {et~al.}(2019)\citenamefont
  {Hussein}, \citenamefont {Governale}, \citenamefont {Kohler}, \citenamefont
  {Belzig}, \citenamefont {Giazotto},\ and\ \citenamefont
  {Braggio}}]{hussein_nonlocal_2019}%
  \BibitemOpen
  \bibfield  {author} {\bibinfo {author} {\bibfnamefont {R.}~\bibnamefont
  {Hussein}}, \bibinfo {author} {\bibfnamefont {M.}~\bibnamefont {Governale}},
  \bibinfo {author} {\bibfnamefont {S.}~\bibnamefont {Kohler}}, \bibinfo
  {author} {\bibfnamefont {W.}~\bibnamefont {Belzig}}, \bibinfo {author}
  {\bibfnamefont {F.}~\bibnamefont {Giazotto}},\ and\ \bibinfo {author}
  {\bibfnamefont {A.}~\bibnamefont {Braggio}},\ }\bibfield  {title} {\bibinfo
  {title} {Nonlocal thermoelectricity in a {Cooper}-pair splitter},\ }\href
  {https://doi.org/10.1103/PhysRevB.99.075429} {\bibfield  {journal} {\bibinfo
  {journal} {Phys. Rev. B}\ }\textbf {\bibinfo {volume} {99}},\ \bibinfo
  {pages} {075429} (\bibinfo {year} {2019})}\BibitemShut {NoStop}%
\bibitem [{\citenamefont {S\'anchez}\ \emph {et~al.}(2019)\citenamefont
  {S\'anchez}, \citenamefont {S\'anchez}, \citenamefont {L\'opez},\ and\
  \citenamefont {Sothmann}}]{sanchez_nonlinear_2019}%
  \BibitemOpen
  \bibfield  {author} {\bibinfo {author} {\bibfnamefont {D.}~\bibnamefont
  {S\'anchez}}, \bibinfo {author} {\bibfnamefont {R.}~\bibnamefont
  {S\'anchez}}, \bibinfo {author} {\bibfnamefont {R.}~\bibnamefont {L\'opez}},\
  and\ \bibinfo {author} {\bibfnamefont {B.}~\bibnamefont {Sothmann}},\
  }\bibfield  {title} {\bibinfo {title} {Nonlinear chiral refrigerators},\
  }\href {https://doi.org/10.1103/PhysRevB.99.245304} {\bibfield  {journal}
  {\bibinfo  {journal} {Phys. Rev. B}\ }\textbf {\bibinfo {volume} {99}},\
  \bibinfo {pages} {245304} (\bibinfo {year} {2019})}\BibitemShut {NoStop}%
\bibitem [{\citenamefont {Dar{\'e}}(2019)}]{dare2019comparative}%
  \BibitemOpen
  \bibfield  {author} {\bibinfo {author} {\bibfnamefont {A.-M.}\ \bibnamefont
  {Dar{\'e}}},\ }\bibfield  {title} {\bibinfo {title} {Comparative study of
  heat-driven and power-driven refrigerators with coulomb-coupled quantum
  dots},\ }\href@noop {} {\bibfield  {journal} {\bibinfo  {journal} {Physical
  Review B}\ }\textbf {\bibinfo {volume} {100}},\ \bibinfo {pages} {195427}
  (\bibinfo {year} {2019})}\BibitemShut {NoStop}%
\bibitem [{\citenamefont {Hwang}\ \emph {et~al.}(2023)\citenamefont {Hwang},
  \citenamefont {Sothmann},\ and\ \citenamefont
  {S\'anchez}}]{hwang_superconductor_2023}%
  \BibitemOpen
  \bibfield  {author} {\bibinfo {author} {\bibfnamefont {S.-Y.}\ \bibnamefont
  {Hwang}}, \bibinfo {author} {\bibfnamefont {B.}~\bibnamefont {Sothmann}},\
  and\ \bibinfo {author} {\bibfnamefont {D.}~\bibnamefont {S\'anchez}},\
  }\bibfield  {title} {\bibinfo {title} {Superconductor--quantum dot hybrid
  coolers},\ }\href {https://doi.org/10.1103/PhysRevB.107.245412} {\bibfield
  {journal} {\bibinfo  {journal} {Phys. Rev. B}\ }\textbf {\bibinfo {volume}
  {107}},\ \bibinfo {pages} {245412} (\bibinfo {year} {2023})}\BibitemShut
  {NoStop}%
\bibitem [{\citenamefont {Kim}\ and\ \citenamefont
  {Hershfield}(2003)}]{kim_thermoelectric_2003}%
  \BibitemOpen
  \bibfield  {author} {\bibinfo {author} {\bibfnamefont {T.-S.}\ \bibnamefont
  {Kim}}\ and\ \bibinfo {author} {\bibfnamefont {S.}~\bibnamefont
  {Hershfield}},\ }\bibfield  {title} {\bibinfo {title} {Thermoelectric effects
  of an {Aharonov}-{Bohm} interferometer with an embedded quantum dot in the
  {Kondo} regime},\ }\href {https://doi.org/10.1103/PhysRevB.67.165313}
  {\bibfield  {journal} {\bibinfo  {journal} {Phys. Rev. B}\ }\textbf {\bibinfo
  {volume} {67}},\ \bibinfo {pages} {165313} (\bibinfo {year}
  {2003})}\BibitemShut {NoStop}%
\bibitem [{\citenamefont {Entin-Wohlman}\ \emph {et~al.}(2010)\citenamefont
  {Entin-Wohlman}, \citenamefont {Imry},\ and\ \citenamefont
  {Aharony}}]{entin-wohlman_three-terminal_2010}%
  \BibitemOpen
  \bibfield  {author} {\bibinfo {author} {\bibfnamefont {O.}~\bibnamefont
  {Entin-Wohlman}}, \bibinfo {author} {\bibfnamefont {Y.}~\bibnamefont
  {Imry}},\ and\ \bibinfo {author} {\bibfnamefont {A.}~\bibnamefont
  {Aharony}},\ }\bibfield  {title} {\bibinfo {title} {Three-terminal
  thermoelectric transport through a molecular junction},\ }\href
  {https://doi.org/10.1103/PhysRevB.82.115314} {\bibfield  {journal} {\bibinfo
  {journal} {Phys. Rev. B}\ }\textbf {\bibinfo {volume} {82}},\ \bibinfo
  {pages} {115314} (\bibinfo {year} {2010})}\BibitemShut {NoStop}%
\bibitem [{\citenamefont {Entin-Wohlman}\ and\ \citenamefont
  {Aharony}(2012)}]{entin-wohlman_three-terminal_2012}%
  \BibitemOpen
  \bibfield  {author} {\bibinfo {author} {\bibfnamefont {O.}~\bibnamefont
  {Entin-Wohlman}}\ and\ \bibinfo {author} {\bibfnamefont {A.}~\bibnamefont
  {Aharony}},\ }\bibfield  {title} {\bibinfo {title} {Three-terminal
  thermoelectric transport under broken time-reversal symmetry},\ }\href
  {https://doi.org/10.1103/PhysRevB.85.085401} {\bibfield  {journal} {\bibinfo
  {journal} {Phys. Rev. B}\ }\textbf {\bibinfo {volume} {85}},\ \bibinfo
  {pages} {085401} (\bibinfo {year} {2012})}\BibitemShut {NoStop}%
\bibitem [{\citenamefont {Hwang}\ \emph {et~al.}(2013)\citenamefont {Hwang},
  \citenamefont {Lim}, \citenamefont {López}, \citenamefont {Lee},\ and\
  \citenamefont {Sánchez}}]{hwang_proposal_2013}%
  \BibitemOpen
  \bibfield  {author} {\bibinfo {author} {\bibfnamefont {S.-Y.}\ \bibnamefont
  {Hwang}}, \bibinfo {author} {\bibfnamefont {J.~S.}\ \bibnamefont {Lim}},
  \bibinfo {author} {\bibfnamefont {R.}~\bibnamefont {López}}, \bibinfo
  {author} {\bibfnamefont {M.}~\bibnamefont {Lee}},\ and\ \bibinfo {author}
  {\bibfnamefont {D.}~\bibnamefont {Sánchez}},\ }\bibfield  {title} {\bibinfo
  {title} {Proposal for a local heating driven spin current generator},\ }\href
  {https://doi.org/10.1063/1.4826108} {\bibfield  {journal} {\bibinfo
  {journal} {Applied Physics Letters}\ }\textbf {\bibinfo {volume} {103}},\
  \bibinfo {pages} {172401} (\bibinfo {year} {2013})}\BibitemShut {NoStop}%
\bibitem [{\citenamefont {Samuelsson}\ \emph {et~al.}(2017)\citenamefont
  {Samuelsson}, \citenamefont {Kheradsoud},\ and\ \citenamefont
  {Sothmann}}]{samuelsson_optimal_2017}%
  \BibitemOpen
  \bibfield  {author} {\bibinfo {author} {\bibfnamefont {P.}~\bibnamefont
  {Samuelsson}}, \bibinfo {author} {\bibfnamefont {S.}~\bibnamefont
  {Kheradsoud}},\ and\ \bibinfo {author} {\bibfnamefont {B.}~\bibnamefont
  {Sothmann}},\ }\bibfield  {title} {\bibinfo {title} {Optimal {Quantum}
  {Interference} {Thermoelectric} {Heat} {Engine} with {Edge} {States}},\
  }\href {https://doi.org/10.1103/PhysRevLett.118.256801} {\bibfield  {journal}
  {\bibinfo  {journal} {Phys. Rev. Lett.}\ }\textbf {\bibinfo {volume} {118}},\
  \bibinfo {pages} {256801} (\bibinfo {year} {2017})}\BibitemShut {NoStop}%
\bibitem [{\citenamefont {Haack}\ and\ \citenamefont
  {Giazotto}(2019)}]{haack_efficient_2019}%
  \BibitemOpen
  \bibfield  {author} {\bibinfo {author} {\bibfnamefont {G.}~\bibnamefont
  {Haack}}\ and\ \bibinfo {author} {\bibfnamefont {F.}~\bibnamefont
  {Giazotto}},\ }\bibfield  {title} {\bibinfo {title} {Efficient and tunable
  {Aharonov}-{Bohm} quantum heat engine},\ }\href
  {https://doi.org/10.1103/PhysRevB.100.235442} {\bibfield  {journal} {\bibinfo
   {journal} {Phys. Rev. B}\ }\textbf {\bibinfo {volume} {100}},\ \bibinfo
  {pages} {235442} (\bibinfo {year} {2019})}\BibitemShut {NoStop}%
\bibitem [{\citenamefont {Blasi}\ \emph {et~al.}(2022)\citenamefont {Blasi},
  \citenamefont {Giazotto},\ and\ \citenamefont {Haack}}]{blasi_hybrid_2023}%
  \BibitemOpen
  \bibfield  {author} {\bibinfo {author} {\bibfnamefont {G.}~\bibnamefont
  {Blasi}}, \bibinfo {author} {\bibfnamefont {F.}~\bibnamefont {Giazotto}},\
  and\ \bibinfo {author} {\bibfnamefont {G.}~\bibnamefont {Haack}},\ }\bibfield
   {title} {\bibinfo {title} {Hybrid normal-superconducting aharonov-bohm
  quantum thermal device},\ }\href {https://doi.org/10.1088/2058-9565/acacbf}
  {\bibfield  {journal} {\bibinfo  {journal} {Quantum Science and Technology}\
  }\textbf {\bibinfo {volume} {8}},\ \bibinfo {pages} {015023} (\bibinfo {year}
  {2022})}\BibitemShut {NoStop}%
\bibitem [{\citenamefont {Yacoby}\ \emph {et~al.}(1995)\citenamefont {Yacoby},
  \citenamefont {Heiblum}, \citenamefont {Mahalu},\ and\ \citenamefont
  {Shtrikman}}]{PhysRevLett.74.4047}%
  \BibitemOpen
  \bibfield  {author} {\bibinfo {author} {\bibfnamefont {A.}~\bibnamefont
  {Yacoby}}, \bibinfo {author} {\bibfnamefont {M.}~\bibnamefont {Heiblum}},
  \bibinfo {author} {\bibfnamefont {D.}~\bibnamefont {Mahalu}},\ and\ \bibinfo
  {author} {\bibfnamefont {H.}~\bibnamefont {Shtrikman}},\ }\bibfield  {title}
  {\bibinfo {title} {Coherence and phase sensitive measurements in a quantum
  dot},\ }\href {https://doi.org/10.1103/PhysRevLett.74.4047} {\bibfield
  {journal} {\bibinfo  {journal} {Phys. Rev. Lett.}\ }\textbf {\bibinfo
  {volume} {74}},\ \bibinfo {pages} {4047} (\bibinfo {year}
  {1995})}\BibitemShut {NoStop}%
\bibitem [{\citenamefont {Keyser}\ \emph {et~al.}(2003)\citenamefont {Keyser},
  \citenamefont {F\"uhner}, \citenamefont {Borck}, \citenamefont {Haug},
  \citenamefont {Bichler}, \citenamefont {Abstreiter},\ and\ \citenamefont
  {Wegscheider}}]{PhysRevLett.90.196601}%
  \BibitemOpen
  \bibfield  {author} {\bibinfo {author} {\bibfnamefont {U.~F.}\ \bibnamefont
  {Keyser}}, \bibinfo {author} {\bibfnamefont {C.}~\bibnamefont {F\"uhner}},
  \bibinfo {author} {\bibfnamefont {S.}~\bibnamefont {Borck}}, \bibinfo
  {author} {\bibfnamefont {R.~J.}\ \bibnamefont {Haug}}, \bibinfo {author}
  {\bibfnamefont {M.}~\bibnamefont {Bichler}}, \bibinfo {author} {\bibfnamefont
  {G.}~\bibnamefont {Abstreiter}},\ and\ \bibinfo {author} {\bibfnamefont
  {W.}~\bibnamefont {Wegscheider}},\ }\bibfield  {title} {\bibinfo {title}
  {Kondo effect in a few-electron quantum ring},\ }\href
  {https://doi.org/10.1103/PhysRevLett.90.196601} {\bibfield  {journal}
  {\bibinfo  {journal} {Phys. Rev. Lett.}\ }\textbf {\bibinfo {volume} {90}},\
  \bibinfo {pages} {196601} (\bibinfo {year} {2003})}\BibitemShut {NoStop}%
\bibitem [{\citenamefont {Aikawa}\ \emph {et~al.}(2004)\citenamefont {Aikawa},
  \citenamefont {Kobayashi}, \citenamefont {Sano}, \citenamefont {Katsumoto},\
  and\ \citenamefont {Iye}}]{PhysRevLett.92.176802}%
  \BibitemOpen
  \bibfield  {author} {\bibinfo {author} {\bibfnamefont {H.}~\bibnamefont
  {Aikawa}}, \bibinfo {author} {\bibfnamefont {K.}~\bibnamefont {Kobayashi}},
  \bibinfo {author} {\bibfnamefont {A.}~\bibnamefont {Sano}}, \bibinfo {author}
  {\bibfnamefont {S.}~\bibnamefont {Katsumoto}},\ and\ \bibinfo {author}
  {\bibfnamefont {Y.}~\bibnamefont {Iye}},\ }\bibfield  {title} {\bibinfo
  {title} {Observation of ``partial coherence'' in an aharonov-bohm
  interferometer with a quantum dot},\ }\href
  {https://doi.org/10.1103/PhysRevLett.92.176802} {\bibfield  {journal}
  {\bibinfo  {journal} {Phys. Rev. Lett.}\ }\textbf {\bibinfo {volume} {92}},\
  \bibinfo {pages} {176802} (\bibinfo {year} {2004})}\BibitemShut {NoStop}%
\bibitem [{\citenamefont {Sigrist}\ \emph {et~al.}(2007)\citenamefont
  {Sigrist}, \citenamefont {Ihn}, \citenamefont {Ensslin}, \citenamefont
  {Reinwald},\ and\ \citenamefont {Wegscheider}}]{PhysRevLett.98.036805}%
  \BibitemOpen
  \bibfield  {author} {\bibinfo {author} {\bibfnamefont {M.}~\bibnamefont
  {Sigrist}}, \bibinfo {author} {\bibfnamefont {T.}~\bibnamefont {Ihn}},
  \bibinfo {author} {\bibfnamefont {K.}~\bibnamefont {Ensslin}}, \bibinfo
  {author} {\bibfnamefont {M.}~\bibnamefont {Reinwald}},\ and\ \bibinfo
  {author} {\bibfnamefont {W.}~\bibnamefont {Wegscheider}},\ }\bibfield
  {title} {\bibinfo {title} {Coherent probing of excited quantum dot states in
  an interferometer},\ }\href {https://doi.org/10.1103/PhysRevLett.98.036805}
  {\bibfield  {journal} {\bibinfo  {journal} {Phys. Rev. Lett.}\ }\textbf
  {\bibinfo {volume} {98}},\ \bibinfo {pages} {036805} (\bibinfo {year}
  {2007})}\BibitemShut {NoStop}%
\bibitem [{\citenamefont {Debbarma}\ \emph {et~al.}(2022)\citenamefont
  {Debbarma}, \citenamefont {Potts}, \citenamefont {Stenberg}, \citenamefont
  {Tsintzis}, \citenamefont {Lehmann}, \citenamefont {Dick}, \citenamefont
  {Leijnse},\ and\ \citenamefont {Thelander}}]{Debbarma2022}%
  \BibitemOpen
  \bibfield  {author} {\bibinfo {author} {\bibfnamefont {R.}~\bibnamefont
  {Debbarma}}, \bibinfo {author} {\bibfnamefont {H.}~\bibnamefont {Potts}},
  \bibinfo {author} {\bibfnamefont {C.~J.}\ \bibnamefont {Stenberg}}, \bibinfo
  {author} {\bibfnamefont {A.}~\bibnamefont {Tsintzis}}, \bibinfo {author}
  {\bibfnamefont {S.}~\bibnamefont {Lehmann}}, \bibinfo {author} {\bibfnamefont
  {K.}~\bibnamefont {Dick}}, \bibinfo {author} {\bibfnamefont {M.}~\bibnamefont
  {Leijnse}},\ and\ \bibinfo {author} {\bibfnamefont {C.}~\bibnamefont
  {Thelander}},\ }\bibfield  {title} {\bibinfo {title} {Effects of parity and
  symmetry on the aharonov--bohm phase of a quantum ring},\ }\href
  {https://doi.org/10.1021/acs.nanolett.1c03882} {\bibfield  {journal}
  {\bibinfo  {journal} {Nano Letters}\ }\textbf {\bibinfo {volume} {22}},\
  \bibinfo {pages} {334} (\bibinfo {year} {2022})}\BibitemShut {NoStop}%
\bibitem [{\citenamefont {Sun}\ \emph {et~al.}(2005)\citenamefont {Sun},
  \citenamefont {Wang},\ and\ \citenamefont {Guo}}]{PhysRevB.71.165310}%
  \BibitemOpen
  \bibfield  {author} {\bibinfo {author} {\bibfnamefont {Q.-f.}\ \bibnamefont
  {Sun}}, \bibinfo {author} {\bibfnamefont {J.}~\bibnamefont {Wang}},\ and\
  \bibinfo {author} {\bibfnamefont {H.}~\bibnamefont {Guo}},\ }\bibfield
  {title} {\bibinfo {title} {Quantum transport theory for nanostructures with
  rashba spin-orbital interaction},\ }\href
  {https://doi.org/10.1103/PhysRevB.71.165310} {\bibfield  {journal} {\bibinfo
  {journal} {Phys. Rev. B}\ }\textbf {\bibinfo {volume} {71}},\ \bibinfo
  {pages} {165310} (\bibinfo {year} {2005})}\BibitemShut {NoStop}%
\bibitem [{\citenamefont {Sun}\ and\ \citenamefont
  {Xie}(2006)}]{sun_bias_2006}%
  \BibitemOpen
  \bibfield  {author} {\bibinfo {author} {\bibfnamefont {Q.-f.}\ \bibnamefont
  {Sun}}\ and\ \bibinfo {author} {\bibfnamefont {X.~C.}\ \bibnamefont {Xie}},\
  }\bibfield  {title} {\bibinfo {title} {Bias-controllable intrinsic spin
  polarization in a quantum dot: Proposed scheme based on spin-orbit
  interaction},\ }\href {https://doi.org/10.1103/PhysRevB.73.235301} {\bibfield
   {journal} {\bibinfo  {journal} {Phys. Rev. B}\ }\textbf {\bibinfo {volume}
  {73}},\ \bibinfo {pages} {235301} (\bibinfo {year} {2006})}\BibitemShut
  {NoStop}%
\bibitem [{\citenamefont {Vernek}\ \emph {et~al.}(2009)\citenamefont {Vernek},
  \citenamefont {Sandler},\ and\ \citenamefont {Ulloa}}]{vernek_kondo_2009}%
  \BibitemOpen
  \bibfield  {author} {\bibinfo {author} {\bibfnamefont {E.}~\bibnamefont
  {Vernek}}, \bibinfo {author} {\bibfnamefont {N.}~\bibnamefont {Sandler}},\
  and\ \bibinfo {author} {\bibfnamefont {S.~E.}\ \bibnamefont {Ulloa}},\
  }\bibfield  {title} {\bibinfo {title} {Kondo screening suppression by
  spin-orbit interaction in quantum dots},\ }\href
  {https://doi.org/10.1103/PhysRevB.80.041302} {\bibfield  {journal} {\bibinfo
  {journal} {Phys. Rev. B}\ }\textbf {\bibinfo {volume} {80}},\ \bibinfo
  {pages} {041302} (\bibinfo {year} {2009})}\BibitemShut {NoStop}%
\bibitem [{\citenamefont {Lim}\ \emph {et~al.}(2010{\natexlab{a}})\citenamefont
  {Lim}, \citenamefont {Crisan}, \citenamefont {Sánchez}, \citenamefont
  {López},\ and\ \citenamefont {Grosu}}]{lim_kondo_2010}%
  \BibitemOpen
  \bibfield  {author} {\bibinfo {author} {\bibfnamefont {J.~S.}\ \bibnamefont
  {Lim}}, \bibinfo {author} {\bibfnamefont {M.}~\bibnamefont {Crisan}},
  \bibinfo {author} {\bibfnamefont {D.}~\bibnamefont {Sánchez}}, \bibinfo
  {author} {\bibfnamefont {R.}~\bibnamefont {López}},\ and\ \bibinfo {author}
  {\bibfnamefont {I.}~\bibnamefont {Grosu}},\ }\bibfield  {title} {\bibinfo
  {title} {Kondo effect in spin-orbit mesoscopic interferometers},\ }\href
  {https://doi.org/10.1103/PhysRevB.81.235309} {\bibfield  {journal} {\bibinfo
  {journal} {Phys. Rev. B}\ }\textbf {\bibinfo {volume} {81}},\ \bibinfo
  {pages} {235309} (\bibinfo {year} {2010}{\natexlab{a}})}\BibitemShut
  {NoStop}%
\bibitem [{\citenamefont {Wellstood}\ \emph {et~al.}(1994)\citenamefont
  {Wellstood}, \citenamefont {Urbina},\ and\ \citenamefont
  {Clarke}}]{wellstood_hot_1994}%
  \BibitemOpen
  \bibfield  {author} {\bibinfo {author} {\bibfnamefont {F.~C.}\ \bibnamefont
  {Wellstood}}, \bibinfo {author} {\bibfnamefont {C.}~\bibnamefont {Urbina}},\
  and\ \bibinfo {author} {\bibfnamefont {J.}~\bibnamefont {Clarke}},\
  }\bibfield  {title} {\bibinfo {title} {Hot-electron effects in metals},\
  }\href {https://doi.org/10.1103/PhysRevB.49.5942} {\bibfield  {journal}
  {\bibinfo  {journal} {Phys. Rev. B}\ }\textbf {\bibinfo {volume} {49}},\
  \bibinfo {pages} {5942} (\bibinfo {year} {1994})}\BibitemShut {NoStop}%
\bibitem [{\citenamefont {Park}\ \emph {et~al.}(2000)\citenamefont {Park},
  \citenamefont {Park}, \citenamefont {Lim}, \citenamefont {Anderson},
  \citenamefont {Alivisatos},\ and\ \citenamefont
  {McEuen}}]{park_nanomechanical_2000}%
  \BibitemOpen
  \bibfield  {author} {\bibinfo {author} {\bibfnamefont {H.}~\bibnamefont
  {Park}}, \bibinfo {author} {\bibfnamefont {J.}~\bibnamefont {Park}}, \bibinfo
  {author} {\bibfnamefont {A.~K.~L.}\ \bibnamefont {Lim}}, \bibinfo {author}
  {\bibfnamefont {E.~H.}\ \bibnamefont {Anderson}}, \bibinfo {author}
  {\bibfnamefont {A.~P.}\ \bibnamefont {Alivisatos}},\ and\ \bibinfo {author}
  {\bibfnamefont {P.~L.}\ \bibnamefont {McEuen}},\ }\bibfield  {title}
  {\bibinfo {title} {Nanomechanical oscillations in a single-{C60}
  transistor},\ }\href {https://doi.org/10.1038/35024031} {\bibfield  {journal}
  {\bibinfo  {journal} {Nature}\ }\textbf {\bibinfo {volume} {407}},\ \bibinfo
  {pages} {57} (\bibinfo {year} {2000})}\BibitemShut {NoStop}%
\bibitem [{\citenamefont {Leturcq}\ \emph {et~al.}(2009)\citenamefont
  {Leturcq}, \citenamefont {Stampfer}, \citenamefont {Inderbitzin},
  \citenamefont {Durrer}, \citenamefont {Hierold}, \citenamefont {Mariani},
  \citenamefont {Schultz}, \citenamefont {von Oppen},\ and\ \citenamefont
  {Ensslin}}]{leturcq_franck-condon_2009}%
  \BibitemOpen
  \bibfield  {author} {\bibinfo {author} {\bibfnamefont {R.}~\bibnamefont
  {Leturcq}}, \bibinfo {author} {\bibfnamefont {C.}~\bibnamefont {Stampfer}},
  \bibinfo {author} {\bibfnamefont {K.}~\bibnamefont {Inderbitzin}}, \bibinfo
  {author} {\bibfnamefont {L.}~\bibnamefont {Durrer}}, \bibinfo {author}
  {\bibfnamefont {C.}~\bibnamefont {Hierold}}, \bibinfo {author} {\bibfnamefont
  {E.}~\bibnamefont {Mariani}}, \bibinfo {author} {\bibfnamefont {M.~G.}\
  \bibnamefont {Schultz}}, \bibinfo {author} {\bibfnamefont {F.}~\bibnamefont
  {von Oppen}},\ and\ \bibinfo {author} {\bibfnamefont {K.}~\bibnamefont
  {Ensslin}},\ }\bibfield  {title} {\bibinfo {title} {Franck-{Condon} blockade
  in suspended carbon nanotube quantum dots},\ }\href
  {https://doi.org/10.1038/nphys1234} {\bibfield  {journal} {\bibinfo
  {journal} {Nat Phys}\ }\textbf {\bibinfo {volume} {5}},\ \bibinfo {pages}
  {327} (\bibinfo {year} {2009})}\BibitemShut {NoStop}%
\bibitem [{\citenamefont {Minot}\ \emph {et~al.}(2004)\citenamefont {Minot},
  \citenamefont {Yaish}, \citenamefont {Sazonova},\ and\ \citenamefont
  {McEuen}}]{Minot2004}%
  \BibitemOpen
  \bibfield  {author} {\bibinfo {author} {\bibfnamefont {E.~D.}\ \bibnamefont
  {Minot}}, \bibinfo {author} {\bibfnamefont {Y.}~\bibnamefont {Yaish}},
  \bibinfo {author} {\bibfnamefont {V.}~\bibnamefont {Sazonova}},\ and\
  \bibinfo {author} {\bibfnamefont {P.~L.}\ \bibnamefont {McEuen}},\ }\bibfield
   {title} {\bibinfo {title} {Determination of electron orbital magnetic
  moments in carbon nanotubes},\ }\href {https://doi.org/10.1038/nature02425}
  {\bibfield  {journal} {\bibinfo  {journal} {Nature}\ }\textbf {\bibinfo
  {volume} {428}},\ \bibinfo {pages} {536} (\bibinfo {year}
  {2004})}\BibitemShut {NoStop}%
\bibitem [{\citenamefont {Haug}\ and\ \citenamefont
  {Jauho}(2008)}]{haug_quantum_2008}%
  \BibitemOpen
  \bibfield  {author} {\bibinfo {author} {\bibfnamefont {H.~J.~W.}\
  \bibnamefont {Haug}}\ and\ \bibinfo {author} {\bibfnamefont {A.-P.}\
  \bibnamefont {Jauho}},\ }\href@noop {} {\emph {\bibinfo {title} {Quantum
  Kinetics in Transport and Optics of Semiconductors}}},\ Springer Series in
  Solid-State Sciences, Vol. 123\ (\bibinfo  {publisher} {Springer-Verlag,
  Berlin},\ \bibinfo {year} {2008})\BibitemShut {NoStop}%
\bibitem [{\citenamefont {Lim}\ \emph {et~al.}(2010{\natexlab{b}})\citenamefont
  {Lim}, \citenamefont {S\'anchez},\ and\ \citenamefont
  {L\'opez}}]{PhysRevB.81.155323}%
  \BibitemOpen
  \bibfield  {author} {\bibinfo {author} {\bibfnamefont {J.~S.}\ \bibnamefont
  {Lim}}, \bibinfo {author} {\bibfnamefont {D.}~\bibnamefont {S\'anchez}},\
  and\ \bibinfo {author} {\bibfnamefont {R.}~\bibnamefont {L\'opez}},\
  }\bibfield  {title} {\bibinfo {title} {Magnetoasymmetric transport in a
  mesoscopic interferometer: From the weak to the strong coupling regime},\
  }\href {https://doi.org/10.1103/PhysRevB.81.155323} {\bibfield  {journal}
  {\bibinfo  {journal} {Phys. Rev. B}\ }\textbf {\bibinfo {volume} {81}},\
  \bibinfo {pages} {155323} (\bibinfo {year} {2010}{\natexlab{b}})}\BibitemShut
  {NoStop}%
\bibitem [{\citenamefont {Agarwalla}\ and\ \citenamefont
  {Segal}(2018)}]{agarwalla_assessing_2018}%
  \BibitemOpen
  \bibfield  {author} {\bibinfo {author} {\bibfnamefont {B.~K.}\ \bibnamefont
  {Agarwalla}}\ and\ \bibinfo {author} {\bibfnamefont {D.}~\bibnamefont
  {Segal}},\ }\bibfield  {title} {\bibinfo {title} {Assessing the validity of
  the thermodynamic uncertainty relation in quantum systems},\ }\href
  {https://doi.org/10.1103/PhysRevB.98.155438} {\bibfield  {journal} {\bibinfo
  {journal} {Phys. Rev. B}\ }\textbf {\bibinfo {volume} {98}},\ \bibinfo
  {pages} {1554} (\bibinfo {year} {2018})}\BibitemShut {NoStop}%
\bibitem [{\citenamefont {Kheradsoud}\ \emph {et~al.}(2019)\citenamefont
  {Kheradsoud}, \citenamefont {Dashti}, \citenamefont {Misiorny}, \citenamefont
  {Potts}, \citenamefont {Splettstoesser},\ and\ \citenamefont
  {Samuelsson}}]{kheradsoud_power_2019}%
  \BibitemOpen
  \bibfield  {author} {\bibinfo {author} {\bibfnamefont {S.}~\bibnamefont
  {Kheradsoud}}, \bibinfo {author} {\bibfnamefont {N.}~\bibnamefont {Dashti}},
  \bibinfo {author} {\bibfnamefont {M.}~\bibnamefont {Misiorny}}, \bibinfo
  {author} {\bibfnamefont {P.~P.}\ \bibnamefont {Potts}}, \bibinfo {author}
  {\bibfnamefont {J.}~\bibnamefont {Splettstoesser}},\ and\ \bibinfo {author}
  {\bibfnamefont {P.}~\bibnamefont {Samuelsson}},\ }\bibfield  {title}
  {\bibinfo {title} {Power, efficiency and fluctuations in a quantum point
  contact as steady-state thermoelectric heat engine},\ }\bibfield  {journal}
  {\bibinfo  {journal} {Entropy}\ }\textbf {\bibinfo {volume} {21}},\ \href
  {https://doi.org/10.3390/e21080777} {10.3390/e21080777} (\bibinfo {year}
  {2019})\BibitemShut {NoStop}%
\end{thebibliography}%
\end{document}